\documentclass{jfm_arxiv}

\usepackage{graphicx}
\usepackage{newtxtext}
\usepackage{newtxmath}
\usepackage{natbib}
\usepackage{hyperref}
\hypersetup{
    colorlinks = true,
    urlcolor   = blue,
    citecolor  = black,
    linkcolor  = blue,
}

\newcommand{\RomanNumeralCaps}[1]
\linenumbers

\usepackage{amsmath}
\usepackage{siunitx}
\usepackage{amssymb} %Weitere mathematiscche Symbole
\usepackage{subcaption}

\usepackage{color}

\renewcommand\vec{\mathbf}
\usepackage{bm}
\newcommand{\matr}[1]{\bm{#1}}     % ISO complying version of matrix

\newcommand{\xs}{\mathbf{x}_s}
\newcommand{\zc}{\hat{\zeta}}
\newcommand{\bk}{\mathbf{k}}

\newcommand{\xsp}{\mathbf{x^\prime}_s}
\newcommand{\zp}{z^\prime}
\newcommand{\tp}{t^\prime}

\newcommand{\nablap}{\nabla^\prime}

\newcommand{\zetap}{\zeta^\prime}

\newcommand{\us}{\mathbf{u}_s}
\newcommand{\up}{\mathbf{u}^\prime}
\newcommand{\vel}{\mathbf{u}}
\newcommand{\Pp}{P^\prime}
\newcommand{\Ma}{\text{Ma}}
\newcommand{\Ca}{\text{Ca}}
\newcommand{\Ga}{\text{Ga}}
\renewcommand{\Re}{\text{Re}_\omega}

\title{The spatial structure of electrostatically forced Faraday waves}

\author{S. Dehe\aff{1},
  M. Hartmann\aff{1},
  A. Bandopadhyay\aff{2}
  \and S. Hardt\aff{1}
  \corresp{\email{hardt@nmf.tu-darmstadt.de}}
 }

\affiliation{\aff{1}Fachgebiet Nano- und Mikrofluidik, Fachbereich Maschinenbau, TU Darmstadt, Darmstadt, Germany
\aff{2}Department of Mechanical Engineering, Indian Institute of Technology Kharagpur, Kharagpur, India}

\begin{document}

\maketitle

\begin{abstract}
The instability of the interface between a dielectric and a conducting liquid, excited by a spatially homogeneous interface-normal time-periodic electric field, is studied based on experiments and theory. Special attention is paid to the spatial structure of the excited Faraday waves. The dominant modes of the instability are extracted using high-speed imaging in combination with an algorithm evaluating light refraction at the liquid-liquid interface.
The influence of the liquid viscosities on the critical voltage corresponding to the onset of instability and on the dominant wavelength is studied. Overall, good agreement with theoretical predictions that are based on viscous fluids in an infinite domain is demonstrated.
Depending on the relative influence of the domain boundary, the patterns exhibit either discrete modes corresponding to surface harmonics or boundary-independent patterns. The agreement between experiments and theory confirms that the electrostatically forced Faraday instability is sufficiently well understood, which may pave the way to control electrostatically driven instabilities. Last but not least, the analogies to classical Faraday instabilities may enable new approaches to study effects that have so far only been observed for mechanical forcing.
\end{abstract}

%\begin{keywords}
%Do not enter any keywords here, since this is done during submission.
%\end{keywords}

\section{Introduction}
\label{sec:intro}

The first systematic investigation of the parametric excitation of a liquid layer on top of a vibrating plate was reported by \citet{Faraday1831}, wherein the free fluid interface exhibits a variety of patterns.
\citeauthor{Faraday1831} used several techniques to visualize the interfacial patterns, including the admixture of tracer particles to the fluid, investigating light reflections, as well as the correlation of light absorption in a dyed liquid to the film thickness.
The dominant pattern showed a dependency on the excitation frequency as well as the amplitude, and \citeauthor{Faraday1831} reported the interface to oscillate with half the excitation frequency. 
The latter observation was challenged by \citet{Matthiessen1868, Matthiessen1870}, who observed oscillations of the surface patterns with the same frequency as the actuation (isochronous).
In an attempt to clarify the contradictory results, \citet{Rayleigh1883} repeated the original experiments by \citeauthor{Faraday1831} and determined the interface to oscillate with half the driving frequency.
The theoretical description of the instability of a liquid layer under harmonic oscillation was provided by \citet{Benjamin1954} in the limiting case of an ideal fluid, resulting in a Mathieu equation. Depending on the forcing parameters, interface oscillations isochronous with the forcing frequency (harmonic) or with half the forcing frequency (subharmonic) are permitted, providing an explanation of the differing observations. 
Later on, the effects of viscosity were incorporated into the stability analysis by \citet{Kumar1994} utilizing Floquet analysis. While the inviscid theory results in instability even at negligible forcing, viscous effects lead to finite forcing amplitudes required to induce instability. With increasing liquid viscosity, the onset of instability shifts to higher driving amplitudes. 
Comparison to experiments demonstrated good agreement both for the critical forcing amplitude as well as the wavelengths observed for the instability. 

Analogous to the mechanical forcing, electric fields at fluid interfaces can induce instabilities. \citet{Yih1968} analyzed the electrohydrodynamic equivalent to the work of \citeauthor{Benjamin1954}, where an interface between a conducting and a dielectric liquid was exposed to a periodically varying electric field.
The stability analysis of the inviscid fluids resulted in a Mathieu equation, but since the Maxwell stress at the interface is proportional to the square of the electric field, the interface oscillates either isochronous or with twice the frequency of the applied voltage.
Around the same time, other research efforts involving electric fields acting on fluid interfaces were directed towards the response of a liquid layer to a perpendicular DC field \citep{Taylor1965}, as well as the parametric forcing in an AC field \citep{Briskman1968}. 
Further coupling mechanisms between interfacial flow and electric fields were investigated in the context of induced fluidic motion by traveling waves \citep{Melcher1966}, tangential electric fields \citep{Melcher1968} and interface shaping by electric fields \citep{Jones1973}.
\citet{Iino1985} used an electrically induced resonance to determine the surface tension of liquid helium. More recently, \citet{Robinson2000,Robinson2001,Robinson2002} analyzed the actuation of a liquid layer in an AC field within an ozone generator.
\citet{Roberts2009} investigated DC and AC actuation of a thin liquid film in the context of pillaring instabilities and used the AC component as an additional means of control over the pillar dimensions.
The instability of a dielectric-dielectric interface was investigated by \citet{Gambhire2010}, and conductivity effects were specifically addressed by \citet{Gambhire2012}. Also, the instability of the interface between a dielectric and a conducting liquid under AC fields accounting for Debye layer effects was studied by \cite{Gambhire2014}.
\citet{Pillai2018} simulated the nonlinear evolution of the interface between an conductor and dielectric in the long-wavelength limit under an oscillatory electric field, demonstrating that the amplitudes of the Faraday waves saturate, without further growth.

Only recently, the theory of \citet{Yih1968} was extended to incorporate viscous effects, similarly to the extension of \citet{Kumar1994} in the case of the mechanically actuated Faraday instability. First, \citet{Bandopadhyay2017} used Floquet theory to study the stability of a perfect dielectric on top of a perfectly conducting fluid, relating the critical voltage and the instability wavenumber.
The marginal stability curve was obtained for single frequency and multiple frequency forcing. For single frequency harmonic oscillations, the interface oscillates isochronous with the forcing frequency, and upon multi-frequency excitation or additional DC offset, the marginal stability curve shows both harmonic and subharmonic tongues.
\citet{Ward2019} extended the model by relaxing the assumption of perfectly dielectric and perfectly conducting liquids and based their analysis on the leaky-dielectric model. 
Additionally, the voltage corresponding to onset of instability for varying driving frequencies was validated for two experimental test cases, showing good agreement to theory.
Due to the restrictions of the experimental setup, no information could be retrieved about the dominant pattern wavelengths. Instead, it was reported that the experimental domain was sufficiently dimensioned to exclude any finite size effects and that for all observed instability modes, a multitude of wavelengths emerged. The goal of the present paper is to extend the experimental exploration of the electrostatically driven Faraday instability by analyzing the spatial structure of the instability and by determining the pattern wavelengths. Especially the fact that a multitude of wavelengths was observed by \citeauthor{Ward2019} for all experimental conditions above the critical threshold warrants further exploration, since it contrasts observations made for mechanically actuated Faraday waves. Here, usually a dominant pattern wavelength emerges, as will be discussed below. 

While the experimental work on electrically actuated Faraday waves is limited, a large body of work exists on mechanically actuated Faraday waves.
Using inviscid theory, \citet{Benjamin1954} showed that the surface deflection of the Faraday instability can be expanded in a complete orthogonal set of eigenfunctions, which for a circular cylinder of radius $R$ are of the form $S_{l,n} = J_l ( k_{l,n} r ) \, \cos ( l \theta )$, where $J_l$ denotes the Bessel function and $k_{l,n}$ the $n$-th zero of $J^\prime_l (k_{l,n} R )$. These discrete modes exhibit a specific dominant wave number $k_{l,m}$ and the authors used the (2,1) mode for comparison between theory and experiments. Similar modes were observed by \citet{Dodge1965} and showed good agreement to the inviscid theory.
Secondary instabilities at higher driving amplitudes were observed by \citet{Gollub1983}, with successive transitions to spatial disorder of the initially ordered system.
The interactions of different modes with adjacent wavenumbers were investigated by \citet{Ciliberto1984}, where periodic and chaotic fluctuations between modes, as well as their superpositions were observed.
\citet{Douady1988} studied the Faraday instability in a square container, leading to rectangular patterns. Also, the role of the meniscus at the container wall was discussed, since the presence of a sidewall meniscus influences the pattern selection.
In subsequent work \citep{Douady1990}, the influence of the lateral boundary condition was further explored. Here, mensicus effects were suppressed by pinning the meniscus and using a brimful container, resulting in pure modes. Contrary, edge-waves were introduced into the domain when a meniscus with a contact angle different from \SI{90}{\degree} was present, which coupled to the parametric excitation.
The time-average of chaotic patterns was studied by \citet{Gluckman1993}, revealing a highly ordered time-averaged system due to the long-range interactions with the boundaries of a moderately large container.
While previous research attempted to reduce the meniscus effects at the domain boundary by pinning the contact line, \citet{Batson2013} used a different approach, where by proper selection of fluids a wetting film at the wall was established, mimicking a moving contact line.
Thereby, the stress-free boundary condition was mimicked, allowing comparison to theory, without including the effects of contact line dissipation.
Recently, \citet{Shao2021} investigated the mode selection in a brimful container and demonstrated the emergence of the first 50 resonant modes. If a concave meniscus was present, edge waves were introduced, leading to complex mode mixing. 

Given a sufficient size of the container, the Faraday instability exhibits boundary independent patterns. For example, \citet{Tufillaro1989} observed a rectangular pattern upon initial instability formation, which transitioned into a disordered pattern with increasing driving amplitude. 
A variety of patterns were reported by \citet{Edwards1994}, comprising lines, squares, circles, and spirals for the single-frequency forcing, and more complex patterns upon multiple-frequency forcing. Also, the container-size influence was analyzed, demonstrating that both high viscosity and high-frequency driving serve to dampen the edge-waves, rendering the system independent of the boundary.
The corresponding extension of the work of \citet{Kumar1994} to multiple-frequency forcing was performed by \citet{Besson1996}, and validated experimentally. 
In the following, amplitude- and phase-resolved measurements of hexagonal and square structures were performed by \citet{Kityk2005}. 
The role of the frequency shift between the different driving frequencies was analyzed by \citet{Epstein2008}, resulting in a criterion for mode-mixing to occur. 
For additional information on Faraday instabilities, the reader is referred to dedicated reviews on this topic, e.g., \citet{Nevolin1984}, \citet{Miles1990}, \citet{Perlin2000} and \citet{Muller2011}. 

From this short overview over the literature of mechanically excited Faraday waves, it is apparent that both boundary dominated systems as well as boundary independent systems exhibit typical pattern wavelengths, which can even be retained after secondary transitions to chaotic regimes. 
The present work aims to clarify which wavelengths and instability patterns are observed for the electrostatically driven Faraday instability, in order to extend the experiments performed by \citet{Ward2019}. 
Also, we wish to scrutinize the reported emergence of a multitude of wavelengths, without the presence of defined eigenmodes, as this observation differs from the results obtained for mechanical actuation.
The remainder of the paper is structured as follows: 
In section \ref{sec:experimentalDetails}, the experimental setup and the light refraction-based method to measure the pattern wavelength are introduced, jointly with the interface reconstruction algorithm. In section \ref{sec:results}, experimentally obtained critical voltages and pattern wavelengths are presented for varying electrolyte conductivity and liquid viscosity. Furthermore, the spatial structure of the instability is presented, revealing the existence of discrete instability modes and superpositions thereof. Also, the transition to a boundary-independent spatial structure of the instability is observed for higher viscosity and driving frequency.
Finally, we demonstrate the applicability of the theory to mode mixing of two superposed driving frequencies. In section \ref{sec:conclusion}, the results are summarized and discussed. 

\section{Experimental Details and Theoretical Description}
\label{sec:experimentalDetails}

\begin{figure}
  \centerline{\includegraphics[width=\textwidth]{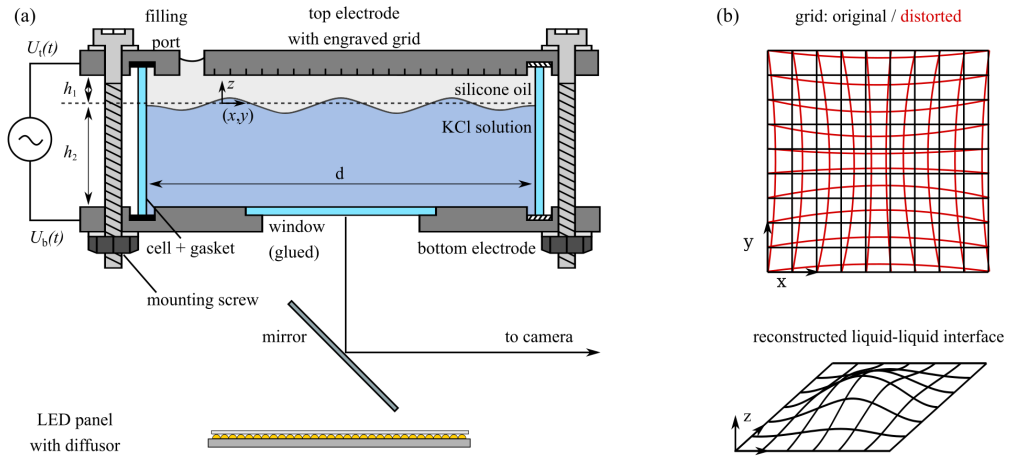}}% Images in 100% size
  \caption{Experimental setup and image processing principle. (a) Schematic of the experimental setup used throughout this study. A high voltage source is connected to the upper and lower electrodes of a cylindrical container with diameter $d$. The container is filled with KCl solution up to a certain height $h_2$. With a layer of silicone oil of height $h_1$ above the KCl solution the container is entirely filled. On the upper electrode surface, a grid is formed by laser engraving. The grid is imaged during an experiment by a high-speed camera via a mirror. Illumination is performed using a LED panel with a diffusor from below. 
  (b) Schematic of the measurement principle. The grid with a line spacing of 0.5\,mm at the upper electrode is imaged and appears distorted due to diffraction at the liquid-liquid interface. From the distorted image, the original interface shape can be reconstructed. 
}
\label{fig:experimentalSetup}
\end{figure}

\subsection{Experimental Setup}

The experimental setup is shown in Figure~\ref{fig:experimentalSetup}(a).
It consists of a cylindrical chamber, a high voltage power source (\textit{HVS448 6000D}, \textit{LabSmith}, USA), a CMOS camera (\textit{FASTCAM Mini AX}, \textit{Photron}, Japan) with a macro objective (\textit{SWM VR ED IF Micro 1:1}, \textit{Nikon}, Japan), a LED panel (\textit{NL480}, \textit{NEEWER}, China), and a mirror cut from a silicon wafer.
The central part of the setup is the cylindrical chamber that is described in the following and represents a modification of the setup used by \citet{Ward2019}.
A circular plate made of stainless steel with a rectangular cut out of $\SI{4.6}{\centi \meter} \times \SI{4.6}{\centi \meter}$ serves as the bottom electrode, leading to an observation region size of approximately $\SI{3.5}{\centi \meter} \times \SI{3.5}{\centi \meter}$.
In order to allow leakage free optical access from below, a rectangular glass window is glued into the cut out using an UV adhesive (\textit{NOA68}, \textit{Thorlabs}, Germany), such that the electrode and the window form a planar connection at the inside of the chamber.
A glass cylinder of height $h \approx h_1 + h_2$ and diameter $d$ allows optical observations from the side.
In the present study $d= \SI{125}{\milli \meter}$, $h_1 = \SI{5}{\milli \meter}$, and $h \approx \SI{35}{\milli \meter}$.
The cylinder is arranged between the bottom and the top electrode by a non-conductive plastic holder, fixated by a set of plastic screws.
Gaskets (\textit{EPDM}, \textit{APSOparts}, Germany) between the glass cylinder and the electrodes assure a leakage free connection.
%Similar to the bottom electrode, the top electrode is a circular stainless steel plate.
A filling port located close to boundary of the cylindrical top plate with a diameter of \SI{1}{\centi \meter} allows filling the chamber with liquids.
At the surface of the upper electrode facing the inside of the chamber, a grating is applied using laser ablation, with a line distance of \SI{0.5}{\milli \meter}.
The grid is imaged through the liquid-liquid interface, and due to diffraction, a distorted image is recorded (see figure \ref{fig:experimentalSetup}(b), red). From the distorted image, the original liquid-liquid interface can be reconstructed, following the procedure outlined by \citet{Moisy2009}, as detailed in section~\ref{sec:dataEvaluation}.
%%The grid with a line distance of \SI{0.5}{\milli \meter} in each direction can be observed in Figure~\ref{fig:experimentalSetup}(b), which is an original image from an experiment.
%%The grid structure is needed to extract the interface deflection (see figure~\ref{fig:experimentalSetup}(c))as a result from a self-written post-processing algorithm as it is described in section~\ref{sec:dataEvaluation}.

To allow optical inspection of the chamber, a rectangular mirror fabricated from a silicon wafer is mounted onto a goniometer (\textit{OWIS}, Germany). It redirects the optical path by \ang{90}, allowing recording by the horizontally oriented high-speed camera via a macro objective.
The goniometer allows fine adjustment of the optical path in two directions.
The illumination of the chamber is performed by a LED panel with a diffusor through the glass window from below, leading to a dark grid appearing on a bright background.
An electric potential difference can be applied between the top and the bottom electrode via connecting each to a separate channel of a high-voltage source. The latter allows generating an AC voltage signal of $\pm \SI{6000}{\volt}$ peak-to-peak, within a sufficiently large frequency range.
At each of the electrodes, a voltage of the form $ U_0/2 \sin \left( 2 \pi f \right)$ with different sign is applied, so that an overall potential difference $U_0\sin \left( 2 \pi f \right)$ is present.
The camera and the high voltage source are both connected to a computer and can be controlled via the appropriate software from the hardware supplier (\textit{Photron Fastcam Viewer} and \textit{Labsmith Sequence}).

\subsection{Experimental Procedure}
\label{sec:experimentalProcedure}

In order to minimize the influence of any pollutant on the liquid and interfacial properties, all parts of the chamber are cleaned thoroughly by rinsing them with isopropanol, followed by DI water.
Then, the parts are blown dry in a nitrogen stream and subsequently assembled to form a leakage free chamber, as sketched in figure~\ref{fig:experimentalSetup}(a).
The chamber is then placed onto a support rack on an air-cushioned optical table, and aligned horizontally.
Prior to the experiments, the optical components are aligned to ensure a distortion-free, focused image of the grid at the upper electrode.
Afterwards, the chamber is filled by first pipetting the bottom liquid through the filling hole up to a level $h_\mathrm{b}$ and then carefully adding a layer of silicone oil above the bottom liquid by the same technique, until it overflows the filling hole.
The liquids used in this study and their properties are described in section~\ref{sec:fluids}.
Before filling, the liquids are degassed within an excesscator for at least 30\,min to avoid gas bubbles in the setup.
The filling port is being covered to avoid dust or other pollutants to enter the chamber after filling.
Subsequently, the electrodes are connected to the two separate channels of the high-voltage source.
Then, the grid is focused by adjusting the focal plane of the objective.\\
\\
In order to measure the critical voltage for a given pair of liquids and a given excitation frequency, the driving signal is chosen with an amplitude well below the theoretically calculated critical voltage.
Then, the excitation is switched on and the system response, i.e., the interface oscillation, is observed for three minutes, both through the camera image and through the side wall of the glass cylinder.
The excitation voltage amplitude is subsequently increased, and the experiment is repeated as described above.
In most cases, it is possible to clearly distinguish between edge-waves stemming from the domain boundary and a critical interface response, i.e., Faraday waves.
If the response of the system exhibits Faraday waves, the image of the grid is recorded for \SI{1}{\second} at a framerate of \SI{1000}{fps} (fps - frames per second), starting \SI{150}{\second} after the voltage has been applied.
Responses without dedicated Faraday patterns are not recorded to reduce data overhead.
In case of an unclear or mixed system response, recordings are made for subsequent clarification via image evaluation.
As described in section~\ref{sec:dataEvaluation}, the data are post-processed in order to identify the type of system response.
In case of strong instabilities, it is necessary to stop an experiment before three minutes have passed in order to prevent the bottom phase to touch the upper electrode, since the deformation amplitude of the interface may increase continuously.
In case of electric breakthrough, liquid of the bottom phase adheres to the upper electrode and influences the electric field in subsequent experiments, thus necessitating a disassembly and cleaning of the chamber.
In case of the interface touching the upper electrode, the data were excluded from the wavelength evaluation, while they were included in the evaluation of the critical voltage.

\subsection{Fluids}
\label{sec:fluids}
\begin{table}
\begin{center}
\def~{\hphantom{0}}
\begin{tabular}{lcccc}
       liquid & density & viscosity & relative permittivity & refractive index\\
        & (\si{\gram\per\milli\litre}) & (\si{\milli\pascal\second}) & (-) & (-) \\[3pt]
       0.65\,cSt	& 0.746 $\pm$ 0.015 & 0.48 $\pm$ 0.01 (c)  & 2.18 (l)& 1.376 \\
       1~~~cSt 		& 0.810 $\pm$ 0.005 & 0.81 $\pm$ 0.01 (c)  & 2.31 (l) & 1.383 \\
       5~~~cSt  	& 0.911 $\pm$ 0.000 & 4.55 $\pm$ 0.00 (c)  & 2.49 (l) & 1.397 \\[3pt]
       0\,~wt\%		& 0.996 $\pm$ 0.010	& 1.00 (l) & - & 1.333	\\
       60\,wt\%		& 1.153	$\pm$ 0.004	& 9.38 $\pm$ 0.43 & - & 1.412	\\
       70\,wt\% 	& 1.185 $\pm$ 0.022 & 19.69 $\pm$ 0.29	& - & 1.427	\\
  \end{tabular}
   \caption{Physical properties of the used liquids. The upper row shows data for the dielectric liquids representing the upper phase, while the bottom row depicts data for the conducting liquids of the lower phase.
Mean values together with standard deviations are reported. In case of viscosity and dielectric permittivity, the values are calculated (c), or taken from the respective data sheet (l).}
  \label{tab:fluidDataExcIFT}
  \end{center}
  \end{table}
  
The conducting lower phase consists of mixtures of DI-water (specific resistance \SI{18.2}{\mega\ohm \centi\meter}, \textit{Milli-Q Integral 3}, \textit{Millipore}) and glycerol (CAS: 56-81-5, Quality: $\geq$ 99.5\,\% water free, \textit{Sigma-Aldrich} and \textit{Carl Roth}, both Germany).
Three different mass fractions of glycerol are used: 0\,wt\%, 60\,wt\% and 70\,wt\%.
Before use, the two components were thoroughly mixed by magnetic stirrer for at least 24\,h to ensure a homogeneous mixture.
A defined amount of potassium chloride (KCl, CAS: 7447-40-7, Quality: ACS Reagent, \textit{Sigma-Aldrich}, Germany) is added to the bottom phase to increase the conductivity.
A concentration of $c_\mathrm{KCl} = \SI{0.001}{\mole\per\liter}$ is used, except for the experiments performed to specifically measure the influence of the concentration.
The fluid data reported in table~\ref{tab:fluidDataExcIFT} and the interfacial tensions in table~\ref{tab:IFTs} were determined with this particular concentration.
For the sake of simplicity, the bottom phase fluids are identified by their glycerol mass fraction as 0\,wt\%, 60\,wt\%, and 70\,wt\%.
Three silicone oils of different viscosity are used (\textit{Silikonöl AK 0.65}, \textit{Silikonöl B1}, and \textit{Silikonöl B5}, \textit{Silikon Profis}, Germany) as dielectric liquids, and hereinafter identified according to their kinematic viscosity as 0.65\,cSt, 1\,cSt, and 5\,cSt.
The fluid properties are summarized in table~\ref{tab:fluidDataExcIFT} and the interfacial tensions in table~\ref{tab:IFTs}.
In Appendix~\ref{sec:appA-substances} it is reported how the fluid properties were measured.

  \begin{table}
  \begin{center}    
        \begin{tabular}{l c c}
         lower phase & upper phase & interfacial tension \\
       & & (\si{\milli\newton\per\metre}) \\[3pt]
        0\,wt\% 	& 0.65\,cSt & 39.6 $\pm$ 0.4 \\
                	& 1~~~cSt & 42.0 $\pm$ 0.1 \\
                	& 5~~~cSt & 35.9 $\pm$ 0.1 \\
        60\,wt\% 	& 0.65\,cSt & 27.4 $\pm$ 0.1 \\        
        70\,wt\% 	& 0.65\,cSt & 25.7 $\pm$ 0.4 	
        \end{tabular}
	\caption{Interfacial tensions between specific fluids. Mean values together with standard deviations are reported.}
		\label{tab:IFTs}
  \end{center}
\end{table}

\subsection{Theoretical description}
\label{sec:theoretical_description}

The theoretical description of the instability is a modification of the approach by \citet{Bandopadhyay2017}, where the upper liquid is considered to be a perfect dielectric, and the lower liquid to be a perfect conductor.
Different from \citet{Bandopadhyay2017}, it is assumed that the upper layer thickness is small compared to that of the lower layer, $h_1/h_2 \ll 1$, in order to simplify the resulting system of equations.
The theoretical approach is described in Appendix~\ref{sec:appB-Theory}, and only the key ideas will be summarized in the following. 
Apart from the work by \citet{Bandopadhyay2017}, a theoretical description exists that takes into account that the dielectric phase can have a non-negligible conductivity \citep{Ward2019}. The authors have shown that the results by \citet{Bandopadhyay2017} are reproduced in the limiting case of an infinite conductivity ratio.
This assumption is specifically revisited in section~\ref{subsec:results_saltvariation}. 

\begin{figure}
	        \centering{
			\includegraphics[width=\textwidth]{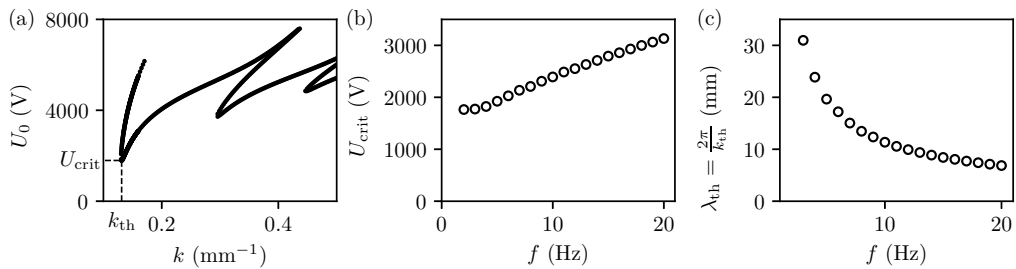}
    \caption{\label{fig:Faraday_Fundamentals_Electrostatic_marginalstability1}Theoretical results obtained for the fluid pair of \SI{0.65}{\centi St} silicone oil and  \mbox{0\,wt\%} glycerol (described in section \ref{sec:fluids}). 
    (a) Marginal stability curve for a driving frequency of $f=\SI{2}{\hertz}$. Different wavenumbers become unstable at different voltages, and the critical voltage $U_\mathrm{crit}$ represents the voltage corresponding to onset of instability considering all wavenumbers.
    (b) Stability map. For otherwise fixed parameters, the critical voltage is plotted versus the driving frequency, and represents the experimentally measurable stability curve. 
    (c) Critical wavelength as a function of critical voltage. Corresponding wavelengths are expected to be observed close to the onset of instability.}
  }
\end{figure}

The system under investigation in \citet{Bandopadhyay2017} consists of two immiscible fluid layers
with thicknesses $h_{1}$, $h_{2}$ between two planar electrodes, as depicted in Fig.~\ref{fig:experimentalSetup}(a), extending infinitely in the lateral direction.
The density $\rho_i$ and the dynamic viscosity $\eta_i$ of both fluids are given, as well as the relative permittivity of the dielectric liquid $\epsilon_1$.
The interface is located at the $z$-position $ \vec{\zeta} (\vec{x}_s,t)$, where $\vec{x}_s$ denotes the position in the $xy$-plane.
The governing equations of the problem are Laplace's equation of electrostatics in the dielectric liquid, and the continuity and Navier-Stokes equations in both liquids. 
The problem is solved using a domain perturbation method for the fluid interface in combination with Floquet theory. Following the mathematical derivation outlined in the Appendix~\ref{sec:appB-Theory}, a generalized eigenvalue problem of the form 
\begin{equation}
\label{eq:eigenvalue_problem}
\matr{A}\vec{Z} = Ma\,\matr{B}\,\vec{Z},
\end{equation}
is obtained, with the matrices $\matr{A}$, $\matr{B}$ and the vector $\vec{Z}$ containing the Fourier-coefficients of $\zeta$. 
The parameter $Ma$ denotes the Mason-number, which is defined as 
\begin{equation}
Ma  = \frac{\epsilon_1 U_0^2}{4 \eta_1 \omega h_1^2},
\end{equation}
where $\omega$ denotes the angular frequency of the driving electric field.
It is important to note that the theoretical description is non-dimensional, however, in the following, the results are shown in dimensionalized form in order to make them more accessible to the reader.
As the problem converges for a finite number of Fourier-modes, as shown by \citet{Bandopadhyay2017}, equation~\ref{eq:eigenvalue_problem} can be used to obtain the marginal stability curve for a given set of fluid pairs, driving frequency and wavenumber $k$.
As shown in the example displayed in figure~\ref{fig:Faraday_Fundamentals_Electrostatic_marginalstability1}(a), the marginal stability curve takes the shape of tongues in the voltage-wavenumber space, characteristic for Faraday instabilities. 
The lowest voltage for a given fluid pair and driving frequency corresponds to the critical voltage $U_\mathrm{crit}$ below which no Faraday instability is expected to occur. 
The corresponding wavenumber $k_\mathrm{th}$ is the most unstable wavenumber, which we expect to observe close to the critical voltage.

Experimentally, the marginal stability curve is not readily measurable, as with increasing voltage amplitude a multitude of wavenumbers become unstable. 
However, the critical voltage $U_\mathrm{crit}$ and the corresponding wavelength $\lambda_\mathrm{th}$ can be measured for different driving frequencies, as shown in figures~\ref{fig:Faraday_Fundamentals_Electrostatic_marginalstability1}(b,c).
For the comparison to experimental observations, both the critical voltage $U_\mathrm{crit}$ and the corresponding wavelength $\lambda_\mathrm{th}$ are used in the following. 
Therefore, we will compute both quantities for varying driving frequencies, with the experimental parameters as input values, and use them to assess the accuracy of the theoretical model.

\subsection{Data Evaluation}
\label{sec:dataEvaluation}

\begin{figure}
  \centerline{\includegraphics[width=\textwidth]{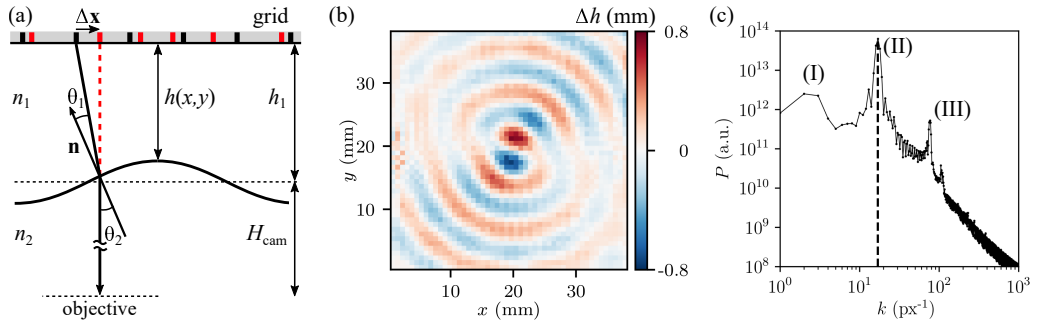}}% Images in 100% size
  \caption{Post-processing steps of the image data. (a) Schematic of the refraction-based image evaluation. A reference grid at the top electrode (black) is imaged through the liquid-liquid interface. Due to different refractive indices of the liquids, the recorded image is distorted (red). (b) Example of a displacement pattern. The displacement field $\Delta \vec{x}$ can be used to determine the surface gradient $\nabla h$, which in turn can be used to reconstruct the interface deflection $\Delta h = h-h_1$. (c) Fourier power spectrum corresponding to (b). The power density is plotted as a function of the wavenumber magnitude $k$ and exhibits several peaks: Peaks (I) and (III) result from the numerical reconstruction algorithm (see text for details). The peak corresponding to the dominant pattern wavelength $\lambda$ (II) is found in an intermediate wavenumber range. Here, $\lambda = 3M\, a_\text{cal} /k = \SI{6.85}{\milli\meter}$.}
\label{fig:experimentalEval}
\end{figure}

For the processing of the experimental data, a refraction-based evaluation routine was used to determine the dominant wavelength of the instability pattern. In Fig. \ref{fig:experimentalEval}(a), a schematic of the measurement principle is shown. The reference grid at the upper electrode (depicted in black) is imaged through the liquid-liquid interface.
Due to the difference of the refractive indices, the image of the grid (depicted in red) is distorted. Light rays passing through horizontal sections of the interface are not altered, whereas rays crossing the interface at positions with a non-zero gradient $\nabla h$ are refracted according to Snell's law ($\sin \theta_1 n_1 = \sin \theta_2 n_2$). The difference between the original grid and the recorded grid is referred to as the displacement field $\Delta \vec{x}$. In the following, we utilize the measurement principle outlined by \citet{Moisy2009}, who derived the relation between the displacement field $\Delta \vec{x}$ and the surface gradient $\nabla h$ introducing the following assumptions: First, paraxial optics is assumed, which requires a small observation field compared to the distance between the objective and the observed grid (here: \SI{3.5}{\centi\meter} vs. \SI{35}{\centi\meter}). Also, they assumed a small slope of the interface, such that the angle between the surface normal $\vec{n}$ and the unit vector in $z$-direction is small, and that the deformation amplitude is small ($\left(h_1-h\right)/h_1 \approx 0$), where $h_1$ is the layer thickness corresponding to the undistorted interface. Then, the relation between the surface gradient $\nabla h$ and the displacement field $\Delta \vec{x}$ is linear, given as 
\begin{equation}
\label{eq:ExpDetails_h_gradient}
\nabla h = -\frac{a_\text{cal}}{h^*} \Delta \vec{x},
\end{equation}
where $a_\text{cal}$ is a calibration factor in units of \SI{}{\milli\meter\per px} obtained from a reference image of a flat interface, and $h^*$ is an effective height obtained from the optical configuration as 
\begin{equation}
\label{eq:ExpDetails_hstar}
h^* = \left( \frac{1}{\alpha h_1}-\frac{1}{H_\text{cam} + h_1} \right)^{-1}. 
\end{equation}
Here, $\alpha = 1-n_2/n_1$ denotes the difference of the refractive indices and $H_\text{cam}$ the distance between the objective and the interface. In our situation, the second term of equation \eqref{eq:ExpDetails_hstar} is negligible due to the large ratio $H_\text{cam}/h_1\ll1$.

Eq. \eqref{eq:ExpDetails_h_gradient} demonstrates that the displacement field for a constant surface gradient $\nabla h$ will increase with increasing layer thickness $h_1$ and a larger difference between the refractive indices $n_1$, $n_2$. While the surface gradients are used to reconstruct the interface $h(x,y)$, it is important to note that we will not rely on the absolute values of the reconstructed interface deformation, but rather on the pattern wavelengths. While the absolute values of $h$ are sensitive to variations in $\nabla h$ as well as $h^*$, the obtained pattern wavelengths. The evaluation scheme includes three distinctive steps: First, the surface gradients $\nabla h(x,y)$ are generated from the displacement field recorded during the experiments.
Second, the interface shape $h(x,y)$ is reconstructed, with one exemplary resulting data set shown in Figure~\ref{fig:experimentalEval}(b).
In a third step, the dominant pattern wavelength is extracted using a Fourier transform and the corresponding power spectrum, as shown in Figure~\ref{fig:experimentalEval}(c).
The details of the evaluation are described in Appendix~\ref{sec:appC-eval}. 

\section{Results and discussions}
\label{sec:results}

In this section, we present the experimental results for the electrically induced Faraday instability.
We focus on the critical voltage and report the dominant pattern wavelengths.
Typically, for one fixed set of experimental parameters, the system shows different characteristic responses with increasing voltage amplitudes. The electric actuation leads to two competing effects: At the side-wall (i.e. the boundary of the container), the meniscus of the dielectric-electrolyte interface oscillates due to the applied Maxwell stress. As the applied Maxwell stress depends quadratically on the electric field strength at the interface, a harmonic driving with a frequency $\omega$ leads to an actuation of the meniscus with a frequency of $2 \omega$. Further, the Faraday instability occurs above a critical voltage, which leads to oscillations with a frequency of $\omega$ when forced with a single frequency. 
At low excitation amplitudes, the actuation of the meniscus leads to small interface deformation with twice the excitation frequency, which is barely noticeable initially. 
With increasing amplitude, the actuation of the meniscus becomes more prominent and waves penetrate from the boundary of the container into the domain. 
A circular pattern is created, with the waves moving into the center of the domain. 
This effect is similar to the edge-wave actuation in the case of mechanically excited Faraday waves.
With further increasing excitation amplitude, Faraday waves start to appear, initially only with a small amplitude. Both edge-waves as well as Faraday waves are present simultaneously. 
A further increase of the amplitude leads to the formation of stronger Faraday waves, which begin to dominate the system. 
Then, a distinct range of driving amplitudes exists, where the pattern becomes stable for long times after an initial growth phase.
A similar phenomenon was observed by \citet{Pillai2018} and attributed to nonlinear effects. 
When some specific excitation amplitude is exceeded, however, the Faraday waves continue to grow until they make contact with the upper electrode and lead to an electric connection between both electrodes.
Then, the power supply shuts off automatically. 
In the following, all parameter combinations that lead to Faraday waves are denoted as critical, since the voltage exceeded the critical voltage.
All other parameters combinations with edge-waves only are denoted as subcritical, since no Faraday patterns emerge during the experiment. 

\subsection{Effect of the salt concentration}
\label{subsec:results_saltvariation}

As we have discussed in section \ref{sec:theoretical_description}, the theoretical model used to predict the instability threshold is based on the perfect-dielectric / perfect-conductor assumption. As was shown by \citet{Ward2019}, the description using the leaky-dielectric model reduces to the perfect-dielectric / perfect-conductor model if the conductivity ratio between the liquids is sufficiently high. In order to assess the validity of the perfect-conductor assumption, we have varied the KCl concentration in the lower liquid (DI-water) from \SI{1e-4}{\mol\per\liter} to \SI{1e-1}{\mol\per\liter} at otherwise fixed parameters (silicone oil viscosity \SI{0.65}{\centi St}, driving frequency \SI{10}{\hertz}).

\begin{figure}
  \centerline{\includegraphics[width=\textwidth]{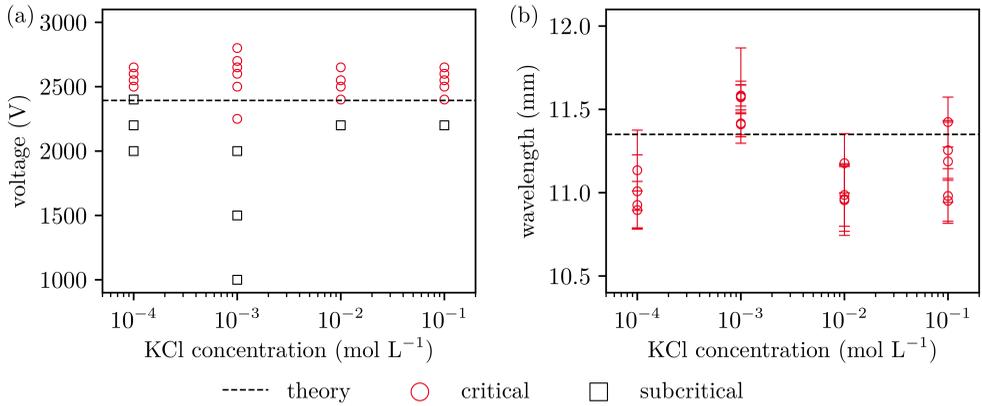}}% Images in 100% size
  \caption{Influence of the lower-phase salt concentration on the instability.
  (a) Experimentally obtained stability map together with the theoretical prediction for the critical voltage obtained from the perfect-dielectric / perfect-conductor model (critical voltage $V_\text{crit}=\SI{2393}{\volt}$).
  (b) Experimentally determined pattern wavelength together with the theoretical prediction for the dominant wavelength obtained from the perfect-dielectric / perfect-conductor model ($\lambda=\SI{11.35}{\milli\meter}$). Each data point corresponds to one critical voltage in subfigure (a). The error bars represent the standard deviation of the obtained wavelength determined within one experiment.}
\label{fig:Results_cSaltStudy}
\end{figure}

In figure \ref{fig:Results_cSaltStudy}(a), the experimentally obtained stability map is shown. As is visible, the critical voltage is not strongly affected by the salt concentration.
It is slightly increased at $c_\text{KCl}=\SI{1e-4}{\mol\per\liter}$ and slightly reduced at $c_\text{KCl}=\SI{1e-3}{\mol\per\liter}$. Since no clearly distinguishable trend is present, we attribute the differences to experimental uncertainties.
One potential source of uncertainty are the edge-waves penetrating into the central region of the system, superposing and obscuring the Faraday waves. Also, since the detection relies on surface gradients, the accuracy at small surface gradients is limited. Nevertheless, the experimentally obtained critical voltages show fair agreement with the theoretically predicted value of $V_\text{crit}=\SI{2393}{\volt}$ (indicated as a dashed line).

In figure \ref{fig:Results_cSaltStudy}(b), the experimentally determined pattern wavelengths corresponding to the critical voltages of figure \ref{fig:Results_cSaltStudy}(a) are shown. As is visible, the change of wavelengths between different driving amplitudes is of the same order of magnitude as the differences over the salt concentration.
The experimentally obtained values show good agreement with the theoretically predicted value of $\lambda =\SI{11.35}{\milli\meter}$. 
Since the experimental results scatter around the theoretically predicted value, the experiments indicate that the salt concentration does not significantly influence the resulting wavelength in the range of concentrations studied. Therefore, the conductivity difference between both liquids is sufficiently large to allow describing the instability using the perfect-dielectric / perfect-conductor model. For the experiments described in the following, we proceed with a constant salt concentration of $c_\text{KCl}=\SI{1e-3}{\mol\per\liter}$.

\subsection{Effect of the viscosity of the dielectric fluid}
In this section, we study the influence of the viscosity of the dielectric fluid by analyzing data from experiments with different silicone oils. While we wish to specifically vary the viscosity, changing the dielectric fluid leads to variations in other parameters as well. A summary of the liquid properties can be found in section \ref{sec:fluids}.
For each driving frequency and voltage amplitude, one experiment was performed. The pattern wavelengths are obtained following the procedure outlined in section \ref{sec:dataEvaluation} and Appendix~\ref{sec:appC-eval}.
For some experiments, no dominant pattern wavelength could be reported , while the corresponding condition is noted as critical in the stability map, i.e., when the interface touched the upper electrode.

\begin{figure}
  \centerline{\includegraphics[width=\textwidth]{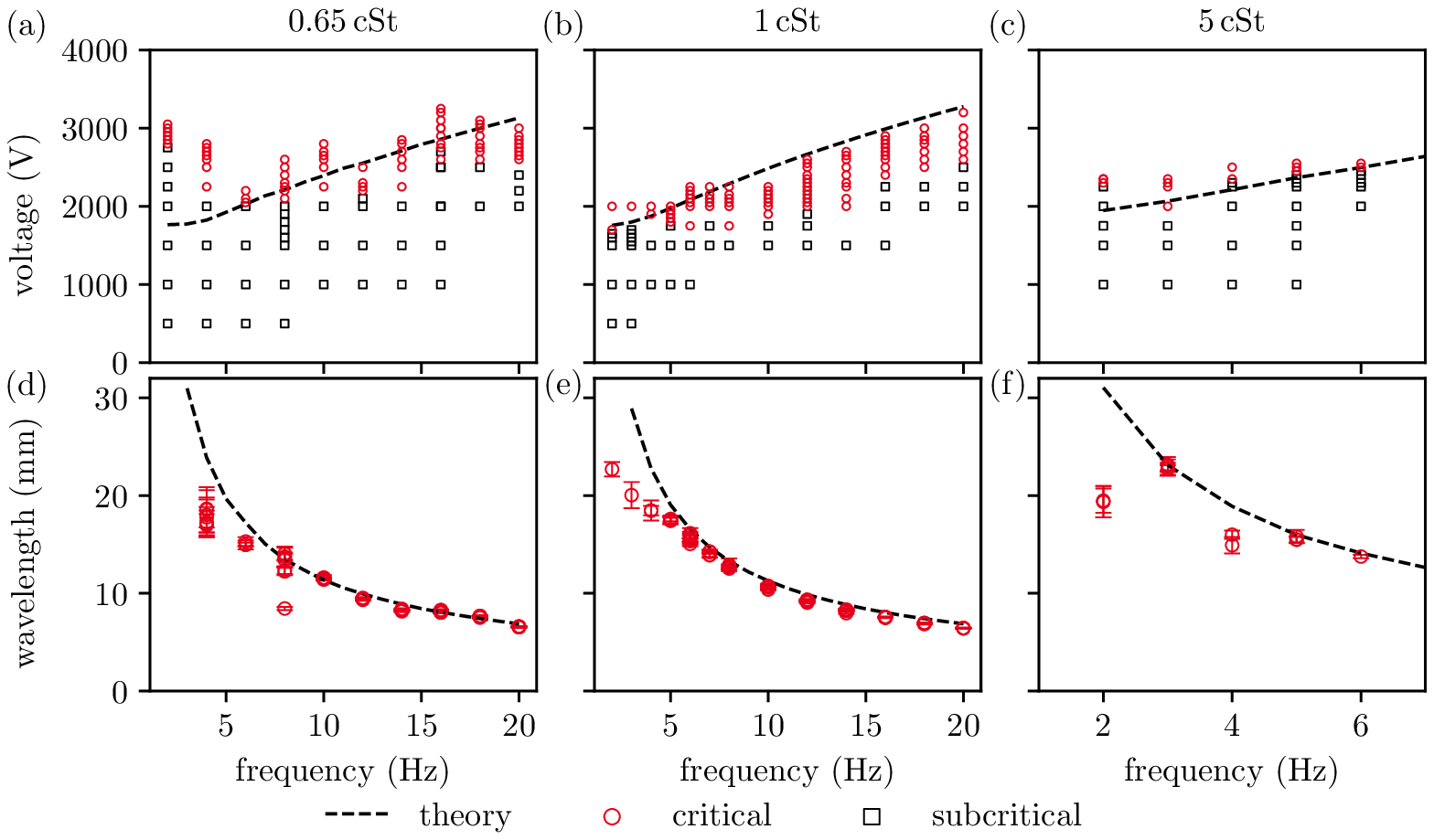}}% Images in 100% size
  \caption{Influence of the viscosity of the dielectric fluid on the instability and the pattern wavelength. DI-water with added KCl of $c=\SI{1}{\mol \per \liter}$ forms the lower phase. 
  (a-c) Experimentally obtained stability maps for different viscosities. 
  (d-f) Dominant wavelengths of the wave patterns for different viscosities. Each data point corresponds to one critical voltage in subfigures (a-c).
 The error bars represent the standard deviation of the obtained wavelength determined within one experiment.
   In all cases, the theoretical predictions are displayed as black lines.}
\label{fig:Results_etaOilStudy}
\end{figure}

In figure \ref{fig:Results_etaOilStudy}, the experimentally obtained stability maps and the resulting pattern wavelengths are displayed, with the theoretical predictions shown as black lines. For the kinematic viscosity of \SI{0.65}{\centi St}, the critical voltage shows partial agreement with the theory. For low excitation frequencies, the onset of the Faraday instability occurs at a significantly higher driving amplitude than predicted. Subsequently, with increasing excitation frequency, the difference between theory and experiments becomes smaller, and at high frequencies, the critical voltage is over-predicted by theory.
The pattern wavelength shows good agreement with theory, and for the multiple critical driving amplitudes at a fixed frequency, similar wavelengths are obtained. 

In Figure~\ref{fig:Results_etaOilStudy}(b,e), results for the silicone oil with a viscosity of \SI{1}{\centi St} are shown. Here, the theory over-predicts the onset of instability compared to the experiments, while a qualitative agreement is apparent. The difference could be explained by an uncertainty of the upper layer thickness. The Maxwell stress at the interface strongly depends on the layer thickness, and a decrease of the thickness of \SI{0.25}{\milli\meter} already increases the Maxwell stress by \SI{10.8}{\percent}. This means that to fix the Maxwell stress with decent accuracy, the layer thickness needs to be determined with very high accuracy.
The pattern wavelength, on the other hand, shows good agreement between theory and experiments, especially for large excitation frequencies. Below \SI{5}{\hertz}, the experimentally obtained wavelengths are smaller than predicted. However, since in that case the wavelengths are of the same scale as the region of interest, the measurement becomes less accurate. Overall, we conclude that there is a quantitative agreement between the experimental and the theoretical data for the pattern wavelength. 

For the oil viscosity of \SI{5}{\centi St}, we report experimental results only up to \SI{6}{\hertz}, as can be seen in figure \ref{fig:Results_etaOilStudy}(c,f). At higher excitation frequencies higher than that, the lower liquid contacted the upper electrode without the occurrence of Faraday waves. Instead, local Taylor cones formed and protruded to the electrode.
For the experimental data we were able to obtain, the agreement of the critical voltage and pattern wavelength is apparent between theory and experiments, with some discrepancies at low frequencies. Again, in that case the pattern wavelength is of the same scale as the observation region, and thus the wavelength might be underpredicted. 

Overall, the experimentally obtained pattern wavelengths and the theoretical predictions agree well. The oil viscosity only has a small influence on the wavelength and the critical voltage. This trend conforms with the theoretical predictions made by \citet{Bandopadhyay2017}.
Also, for a given frequency, a change of driving amplitude above the onset of the Faraday instability has only little effect on the resulting pattern wavelength, since the resulting wavelengths are reproduced for different driving voltages.

\subsection{Effect of the viscosity of the electrolyte}

In this section, we discuss the influence of the lower layer's viscosity on the stability of the system and the wave patterns. For this purpose, glycerol-water mixtures with different glycerol mass fractions, expressed as weight-percent (wt\%), were used. For the electrolyte with \mbox{60\,wt\%} glycerol, the dynamic viscosity is \SI{9.38}{\milli\pascal\second} and for \mbox{70\,wt\%} glycerol it is \SI{19.69}{\milli\pascal\second}.
The upper layer was silicone oil with a kinematic viscosity of \SI{0.65}{\centi St}.
Again, for each excitation frequency and voltage amplitude, one experiment was performed.

\begin{figure}
  \centerline{\includegraphics[width=\textwidth]{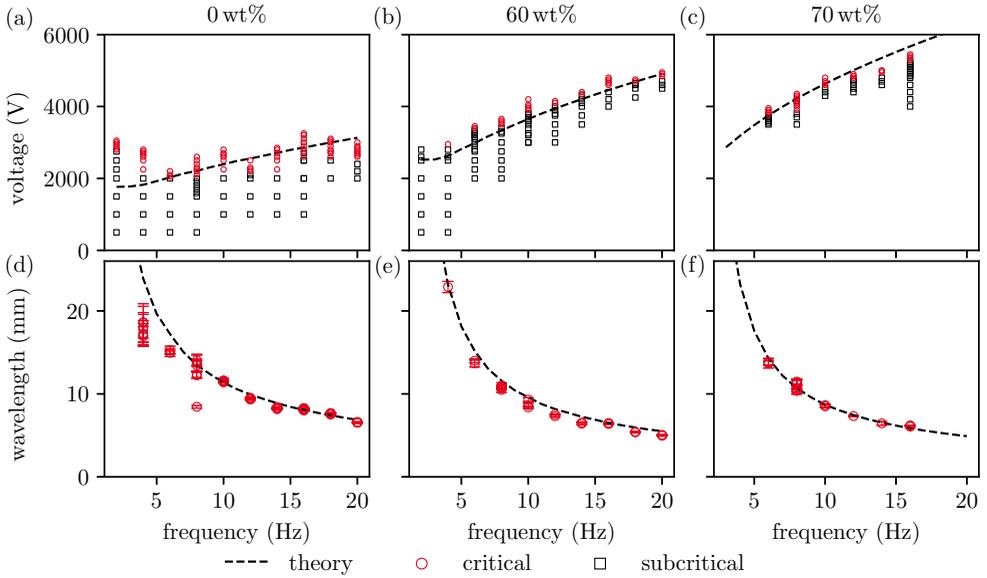}}% Images in 100% size
  \caption{Influence of the viscosity of the dielectric fluid on the instability and the pattern wavelength. Silicone oil with \SI{0.65}{\centi St} forms the upper phase, and the KCl concentration in the lower phase is $c=\SI{1}{\mol \per \liter}$. 
  (a-c) Experimentally obtained stability maps for different electrolytes. 
  (d-f) Dominant wavelengths of the wave patterns for different electrolytes. Each data point corresponds to one critical voltage in subfigures (a-c). The error bars represent the standard deviation of the obtained wavelength determined within one experiment.
  In all cases, the theoretical predictions are displayed as black lines.}
\label{fig:Results_cGlycStudy}
\end{figure}

In figure \ref{fig:Results_cGlycStudy}, the experimentally obtained stability maps and pattern wavelengths are shown, where the data for \mbox{0\,wt\%} glycerol corresponds to figure \ref{fig:Results_etaOilStudy}(a,d).
The data for \mbox{60\,wt\%} glycerol show a good agreement between the experimentally measured and the theoretically predicted critical voltage. 
Compared to the experiments with \mbox{0\,wt\%}, the range of Faraday patterns that do not grow indefinitely is smaller.
Also, compared to the previous experiments, the critical voltage is higher, and its slope increases with the viscosity of the lower phase.
The pattern wavelength is represented well by the theoretical model over the whole range of excitation frequencies, as can be seen from figure \ref{fig:Results_cGlycStudy}(e). 

Increasing the glycerol mass ratio to \mbox{70\,wt\%} leads to even higher critical voltages, as can be seen from figure \ref{fig:Results_cGlycStudy}(c). Here, the agreement between the stability map and the theoretically predicted critical voltage is good as well, with increasing deviations at higher excitation frequencies.
As is visible, the range of Faraday patterns that do not exhibit indefinitely growing amplitudes is even smaller than for \mbox{60\,wt\%}. Often, only one driving voltage exhibited a Faraday pattern that was stable for longer times, and an increase of the driving amplitude lead to a continuously growing interface deflection and ultimately an electric connection between both electrodes. Again, the experimental and the theoretical data for the pattern wavelengths show good agreement.

\subsection{Spatial structure of the wave patterns}

So far, we have focused on the wavelength of the emerging patterns, without further detailing their spatial structure. 
The theoretical considerations presented in section \ref{sec:theoretical_description} assume an infinite domain without lateral boundaries.
While the dominant wavelength and critical voltage are predicted by theory, due to the absence of boundary effects no information about the spatial structure of the waves can be derived from theory.
The experimental results by \citet{Ward2019} were obtained using a fluid domain of similar size and with similar fluid pairs, but due to imaging in side view, the authors could not distinguish specific modes or determine the wavelengths.
They concluded that the domain size was sufficient such that the instability patterns were not influenced by the lateral boundaries, and reported that a multitude of wavelengths are observed during all unstable configurations.
In the following, we report the spatial structure of the observed wave patterns and compare them to the observations made for mechanically excited Faraday waves.

\begin{figure}
  \centerline{\includegraphics[]{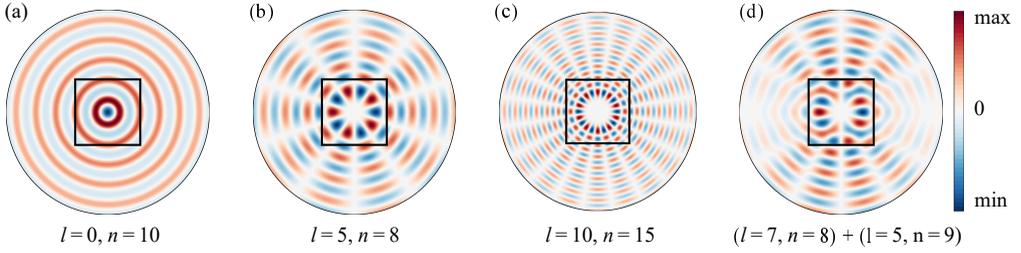}}% Images in 100% size
  \caption{Theoretically predicted Bessel modes of the form $\Delta h = J_l\left( k_n r \right) \cos \left( l \theta \right)$, where $l$ is the azimuthal mode number and $k_n$ are the roots of $J_l^\prime\left( k_n R \right)$. (a-c) Exemplary modes with different mode numbers $l, n$ are shown. The circular region corresponds to the the circular interface of the experiments. Red values corresponds to the peaks of the interface deflection $\Delta h$, and blue to the valleys. Due to the limited field of view, only the central region is observed in our experiments, indicated as black squares. (d) A superposition of two modes creates patterns with odd parity.}
\label{fig:Results_Modes_Theory}
\end{figure}

Before continuing with our experimental results, it is instructive to draw from the analogy to mechanically induced Faraday waves. 
As was shown by \citet{Benjamin1954}, the surface deflection of Faraday waves can be expressed as a series of a complete orthogonal set of eigenfunctions, assuming ideal fluids, taking the form
\begin{equation}
\label{eq:Results_Bessel_Modes}
\Delta h = \sum a_{l,n}(t) J_l\left( k_{l,n} r \right) \cos \left( l \theta \right),
\end{equation}
where $a_{l,n}(t)$ denotes a time-dependent amplitude, $J_l$ the Bessel function with the azimuthal node number $l$, and $k_{l,n}$ the $n$-th root of $J_l^\prime \left( k_{l,n} R \right)$.
In a recent work by \citet{Shao2021}, the modes of the mechanical Faraday instability in a circular container of finite size were characterized, demonstrating similar mode forms. By changing the frequency and the amplitude of the excitation, they were able to observe a large range of pure modes.
In  figure \ref{fig:Results_Modes_Theory}, some exemplary theoretical modes following the form of equation \eqref{eq:Results_Bessel_Modes} are depicted. Apparently, $l$ leads to an increase of the number of nodes in the circumferential direction, and $n$ leads to more nodes in the radial direction. As prominently visible in figures \ref{fig:Results_Modes_Theory}(b,c), a nearly unperturbed region is found at the center of the domain, enclosed by a circular region of larger deflections.
Also, as was discussed for example by \citet{Ciliberto1984}, mode mixing can occur between modes with similar $k_{l,n}$, which can lead to excitation of modes of odd parity with respect to a specific axis, as shown in figure \ref{fig:Results_Modes_Theory}(d).
Owning to the substantial differences between the corresponding systems studied in the literature and our system (air vs. liquid as the upper layer; mechanical vs. electrical actuation), it is not clear \textit{a priori} if the system studied by us will exhibit a dominant boundary influence or if it will behave as an unbounded domain.
In addition, \citet{Shao2021} illuminated the role of the boundary meniscus, leading to waves traveling into the domain, which in the case of mechanical actuation stem from a contact angle different from \SI{90}{\degree} at the container side wall. The superposition of Faraday waves and edge waves leads to complex instability patterns deviating from the modes of equation \ref{eq:Results_Bessel_Modes}. 

In the following, the instability patterns at the moment of largest deflection during an oscillation period are shown. Thus, they depict an instantaneous surface profile and contain no information about the time-evolution or the pattern dynamics. As was revealed by \citet{Gluckman1993} in the context of mechanically actuated Faraday patterns, systems that appear disordered instantaneously can exhibit an ordered time-average over long times.
The interface deflection $\Delta h$ is shown normalized, with the highest and lowest values shown in color (red and blue), such that the spatial structure is visible.
Each image is normalized by assigning red to the maximum and blue to the minimum value present within the image, in order to emphasize the spatial structure.
As we have outlined in section \ref{sec:experimentalProcedure}, the instability is recorded \SI{150}{\second} after the voltage is switched on, such that the patterns are able to reach a quasi-steady state.
For the sake of brevity, not every experimentally obtained interface deformation is shown, but rather representative modes for specific frequencies and amplitudes. Only three combinations of oils and electrolytes are shown, since the other experimental configurations show similar behavior without qualitatively new information.

\begin{figure}
  \centerline{\includegraphics[width=\textwidth]{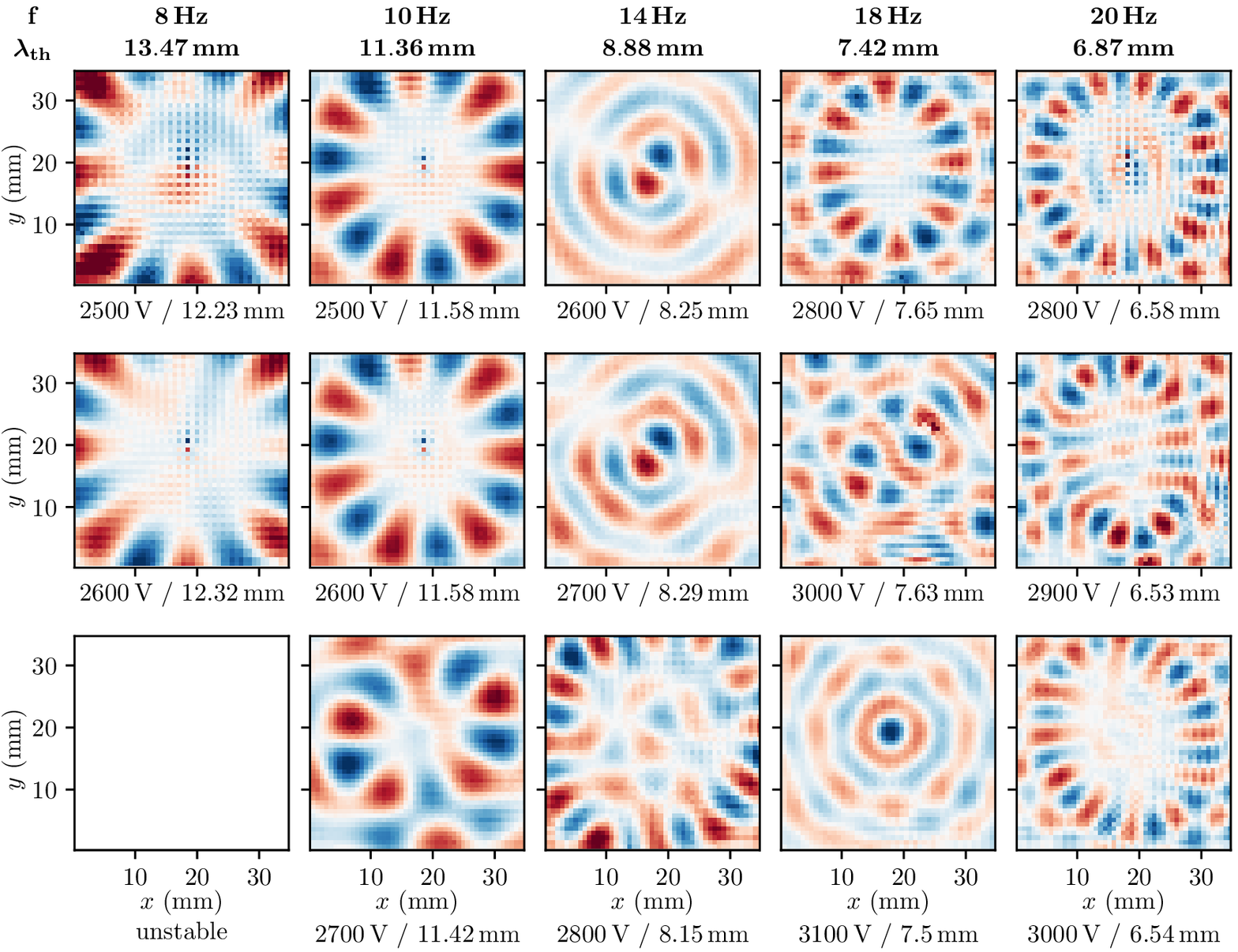}}% Images in 100% size
  \caption{Reconstructed spatial structure of the liquid-liquid interface using silicone oil with a viscosity of \SI{0.65}{\centi St} and DI-water with a salt concentration of $\SI{1}{\milli\mol\per\liter}$. The normalized interface deflection $\Delta h$ is displayed, with blue denoting negative deflections and red positive deflections. At the top of each row, the excitation frequency and the theoretically predicted pattern wavelength are shown. Below each image, the corresponding voltage amplitude and the experimentally obtained pattern wavelength are indicated.}
\label{fig:Results_Modes_0-65cSt_water}
\end{figure}

In figure \ref{fig:Results_Modes_0-65cSt_water}, Faraday patterns for the silicone oil with \SI{0.65}{\centi St} in combination with water are shown, where the excitation frequency and the theoretically predicted wavelength are displayed at the top of each column.
Below each panel, the applied voltage amplitude and the experimentally obtained wavelength are noted and sorted by voltage.
The patterns obtained for \SI{8}{\hertz} and \SI{10}{\hertz} strongly resemble the modes described by \ref{eq:Results_Bessel_Modes}, with $l=8$ and $l=7$, respectively. Both the circular structure and the inner region without distinct patterns are visible. While it is straightforward to identify the azimuthal mode number $l$, the radial mode number $n$ is not easily obtained due to the limited field of view at the center of the container. Also, these results exemplify why the measurements of the wavelength at low driving frequencies show larger deviations from theory. In the central region, the patterns do not display distinct deflections, which complicates the wavelength identification.
Nevertheless, while the observed modes are strongly influenced by the domain boundary, as is shown by the existence of eigenmodes of an interface in a circular container, the experimentally obtained wavelengths correspond to theoretical predictions from the unbound theory reasonably well.
With increasing driving voltage, the instability patterns can change drastically, while maintaining the overall dominant wavelength.
This is for example visible at \SI{10}{\hertz} for the amplitudes of \SI{2600}{\volt} and \SI{2700}{\volt}, where the spatial structure changes and is not well-represented by modes of the form of equation \ref{eq:Results_Bessel_Modes}.
As visible from the patterns at \SI{18}{\hertz} and \SI{3000}{\volt}, also modes without clear boundary influence can be observed. However, as can be seen both for higher and lower voltage amplitudes, at the same frequency also highly ordered spatial structures emerge.
Overall, the Faraday patterns for \SI{0.65}{\centi St} silicone oil in combination with water are mostly dominated by the domain boundary. 

\begin{figure}
  \centerline{\includegraphics[width=\textwidth]{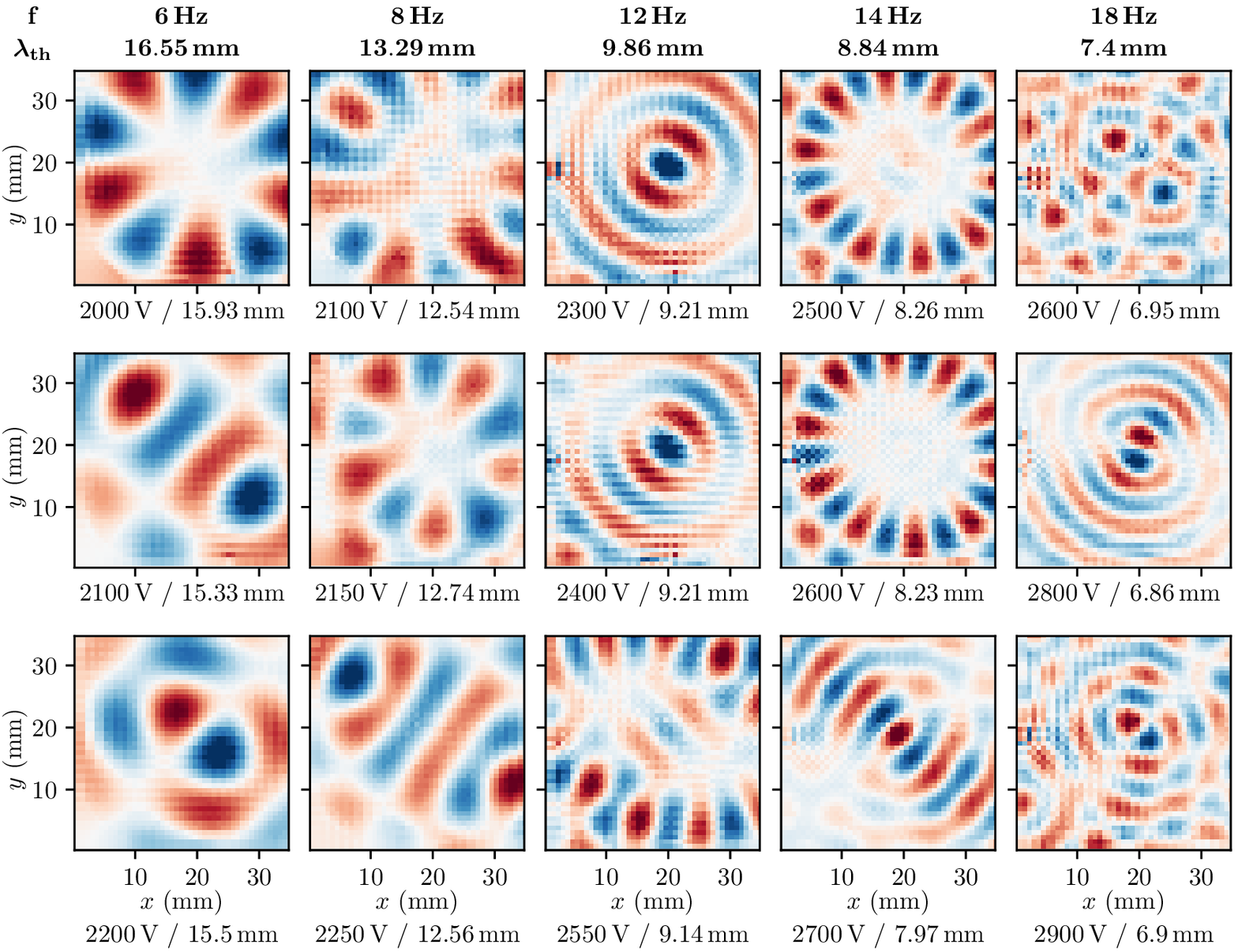}}% Images in 100% size
  \caption{Reconstructed spatial structure of the liquid-liquid interface using silicone oil with a viscosity of \SI{1}{\centi St} and DI-water with a salt concentration of $\SI{1}{\milli\mol\per\liter}$. The normalized interface deflection $\Delta h$ is displayed, with blue denoting negative interface deflections and red positive deflections. At the top of each row, the excitation frequency and the theoretically predicted pattern wavelength are shown. Below each image, the corresponding voltage amplitude and the experimentally obtained pattern wavelength are indicated.
  }
\label{fig:Results_Modes_1cSt_water}
\end{figure}

In figure \ref{fig:Results_Modes_1cSt_water}, Faraday patterns for the silicone oil with \SI{1}{\centi St} in combination with water are shown, with a similar layout as in figure \ref{fig:Results_Modes_0-65cSt_water}.
Here, the wall influence is also prominently visible, yielding modes similar to those predicted by equation \ref{eq:Results_Bessel_Modes}, for example with $l=5$ at \SI{6}{\hertz} and \SI{2000}{\volt}, $l=5$ at \SI{8}{\hertz} and \SI{2150}{\volt}, and $l=10$ at \SI{14}{\hertz} and \SI{2500}{\volt}.
In addition to these pure modes, other highly ordered patterns can be observed. For example, at \SI{8}{\hertz} and \SI{2100}{\volt}, the pattern is anti-symmetric with respect to an axis that is tilted approximately by \SI{30}{\degree} relative to the $y$-axis. A similar pattern is obtained by superposition of the modes ($l_1=7$, $n_1=8$) and ($l_2=5$, $n_2=9$), which can be seen by  comparison to figure \ref{fig:Results_Modes_Theory}(d).
The actuation by an amplitude above the critical voltage can trigger different modes with similar wavelengths, which superposed lead to (anti-)symmetric patterns. Further examples of superpositions of patterns include the patterns at \SI{6}{\hertz} and \SI{2100}{\volt}, \SI{8}{\hertz} and \SI{2250}{\volt} as well as \SI{18}{\hertz} and \SI{2600}{\volt}.
While a systematic classification and decomposition into the eigenmodes of equation \ref{eq:Results_Bessel_Modes} is possible, it is beyond the scope of this work.
In spite of the prominent influence of the boundary, the dominant wavelength extracted from the experiments show good agreement with the theoretically predicted values.
Overall, for this combination of liquids the Faraday patterns are dominated by the domain boundary as well, with the additional occurrence superposed modes. 

\begin{figure}
  \centerline{\includegraphics[width=\textwidth]{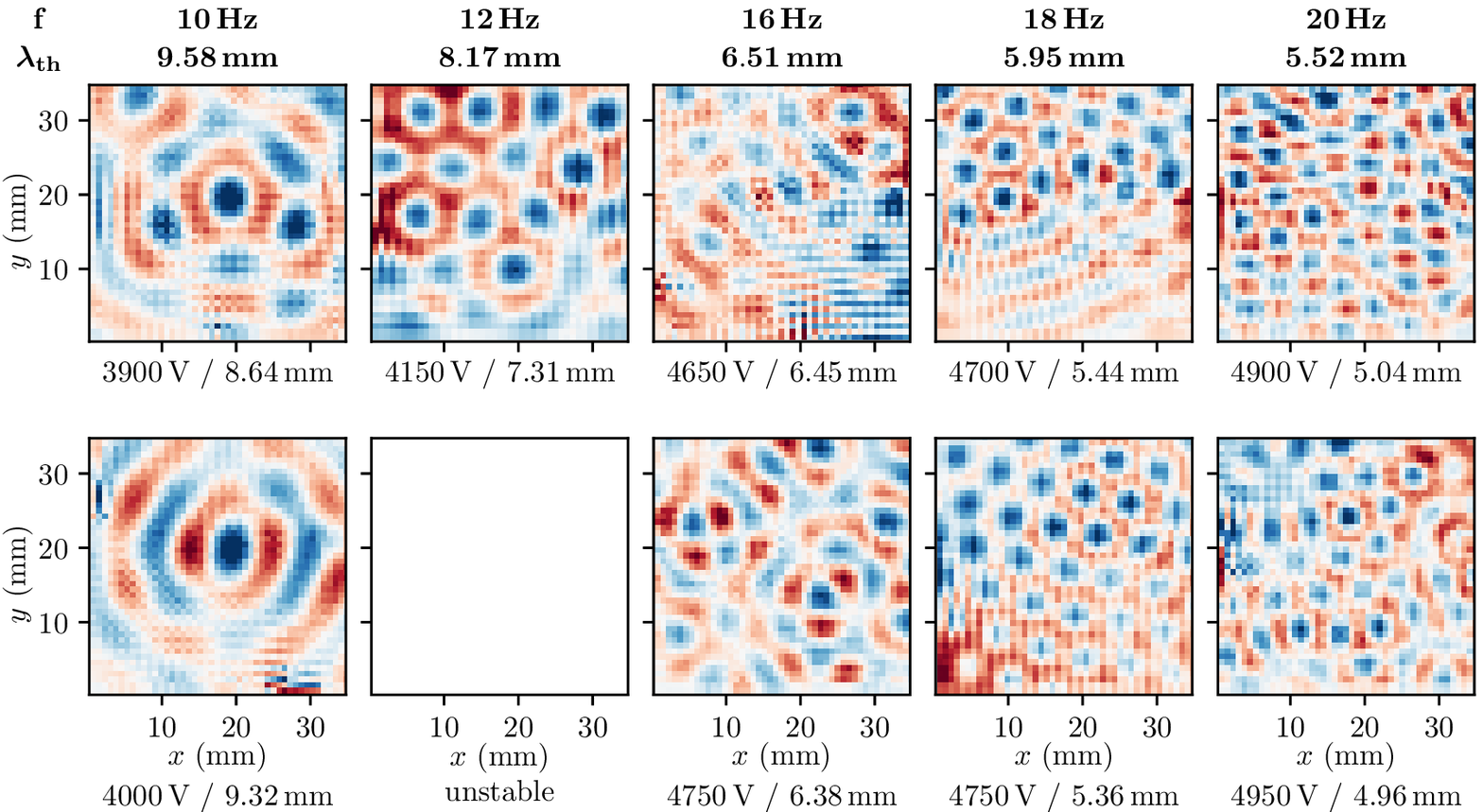}}% Images in 100% size
  \caption{Reconstructed spatial structure of the liquid-liquid interface using silicone oil with a viscosity of \SI{0.65}{\centi St} and a water-glycerol mixture (\mbox{60\,wt\%}) with a salt concentration of $\SI{1}{\milli\mol\per\liter}$. The normalized interface deflection $\Delta h$ is displayed, with blue denoting negative interface deflections and red positive deflections. At the top of each row, the excitation frequency and the theoretically predicted pattern wavelength are shown. Below each image, the corresponding voltage amplitude and the experimentally obtained pattern wavelength are indicated.
  }
\label{fig:Results_Modes_0-65cSt_glyc60}
\end{figure}

The observed Faraday patterns change with increasing viscosity of the lower phase. In figure  \ref{fig:Results_Modes_0-65cSt_glyc60}, the patterns for the silicone oil with \SI{0.65}{\centi St} in combination with a water-glycerol mixture (\mbox{60\,wt\%}) are shown.
Owing to the smaller voltage amplitude range of quasi-stable patterns, only two patterns per excitation frequency are shown.
While at excitation frequencies up to \SI{10}{\hertz} the patterns still resemble the Bessel modes and are thus dominated by the boundary, at higher frequencies the situation changes drastically, and the patterns no longer show a boundary influence.
They resemble those observed for mechanically induced Faraday instabilities \citep{Edwards1994}.
For example, at \SI{12}{\hertz} excitation frequency, the pattern exhibits a typical hexagonal structure, and at \SI{16}{\hertz}, a pattern that is ordered at low voltage amplitudes becomes increasingly chaotic.
At \SI{18}{\hertz} and \SI{20}{\hertz}, the patterns again display hexagonal structures. 
Overall, at higher frequencies and increased viscosity, the boundary effects become negligible. A similar trend was observed by \citet{Edwards1994} for mechanically actuated Faraday waves. The suppression of boundary effects is due to the increased viscosity that leads to strong damping and the comparatively high excitation frequencies that produce wavelengths much smaller than the container size.

At this point, it is instructive to compare the results of this work with the results of \citet{Ward2019}. In both experimental setups, the domain size is comparable, and the thickness of the upper liquid layer is comparable as well. \citeauthor{Ward2019} use oil with a kinematic viscosity of \SI{1.5}{\centi St}, which is most closely matched by the viscosity of \SI{1}{\centi St} in our work. As we have shown, the Faraday patterns are strongly influenced by the domain boundary, and both single eigenmodes as well as superpositions of eigenmodes were observed. At low viscosities, in most cases the patterns are highly ordered, exhibiting characteristic wavelengths corresponding to theory. Only at higher viscosities, the patterns become rather independent of the boundary, especially at higher frequencies. 
However, for both boundary-dominated and boundary-independent patterns, the spatial structures exhibited characteristic wavelengths and were highly organized.
While \citeauthor{Ward2019} did not report specific wavelenghts, they noted that in their experiments the domain could be regarded as effectively unbounded, and that multiple wavelengths were obtained in all experiments.
We hypothesize that two reasons could potentially be responsible for the observed differences: First, in our experiments, the driving voltage was fixed, either leading to an instability or not. In the experiments by \cite{Ward2019}, the instability was induced above the critical voltage, and then the amplitude was reduced successively until the instability was no longer observable. In principle, a multitude of wavelengths might become unstable and coexist well above the critical voltage.
However, as is visible for the instability patterns reported in this section, even experimental conditions well above the first onset of the Faraday instability resulted in well-ordered systems, with characteristic wavelengths. Thus, we would expect similar effects to occur in the experiments by \citeauthor{Ward2019}.
The second explanation relates to the mode of observation.
\citet{Ward2019} observed the interface from the side, whereas in our experiments, the interface is observed from the bottom. As was noted by \citet{Ward2019}, the optical distortion due to refraction of the cylinder, as well as the edge waves obscured the Faraday patterns.
It is likely that the wave pattern was significantly influenced by the domain boundary, leading to discrete modes of the surface harmonics. When imaged from the side, the patterns might have appeared to consist of a multitude of wavelengths. 
Nevertheless, as we have shown in this section, the pattern wavelength in the boundary-dominated case matches the wavelength predicted by the theory for the unbounded case, and similarly, the critical voltage matches the predictions by this theory as well. Thus, the experimental results with respect to the critical voltage by \citeauthor{Ward2019} remain valid.

Last but not least, the results of this section highlight the analogy between mechanical and electrostatic actuation. At high frequencies and viscosities, the Faraday pattern becomes independent of the domain boundary. In addition to the already observed patterns, other patterns should be observable using electrostatic actuation upon proper parameter tuning. For example, different spatial structures, such as rectangular patterns, quasi-patterns \citep{Edwards1994} and superlattices \citep{Douady1990, Kahouadji2015} should be expected for the case that the boundary influence is negligible. They might be interesting to study in the context of applications for the Faraday instability, such as the patterning of thin films \citep{Zhao2019} and patterning of objects suspended in the lower phase, e.g., biological cells \citep{Chen2017}. 

\section{Conclusions}
\label{sec:conclusion}

We have studied the electrostatically forced Faraday instability, placing a focus on the spatial structure of the patterns and their wavelengths. 
For that purpose, the instability was induced at a circular interface between a liquid dielectric and a conducting fluid by applying an oscillatory voltage between two parallel-plate electrodes. 
The interface deformation was reconstructed by imaging a grid through the perturbed interface,. 
From the Fourier power spectrum the dominant wavelength of the instability was extracted and compared to theoretical predictions.

The Faraday patterns were investigated for silicone oils of different viscosities (\SI{0.65}{\centi St} - \SI{5}{\centi St}). The viscosity of the aqueous electrolyte was adjusted by adding glycerol (0\,wt\%-70\,wt\%). 
In order to compare the theoretical predictions of \citet{Bandopadhyay2017} to the experiments, both the critical voltage and the dominant pattern wavelength were extracted for varying excitation frequencies.
The dominant pattern wavelength shows good agreement with the theoretical prediction and was reproducible for a range of voltages above the critical voltage. 
The agreement of the critical voltage between experiments and theory was not satisfactory at low voltage amplitudes but improved at higher amplitudes. 
For the whole range of instabilities observed, the response frequencies of the interface corresponded to the excitation frequency, as expected from theoretical considerations \citep{Bandopadhyay2017, Ward2019}.

For the majority of the experiments, the circular domain boundary had a pronounced influence on the wave patterns. 
A variety of discrete modes were observed, resembling the classical Bessel-function surface harmonics, defined by the pair of azimuthal node number $l$ and radial node number $n$. 
The existence of such modes emphasizes the pronounced role of the domain boundary. 
In analogy to mechanically actuated Faraday waves, the mixing of modes with similar wavelengths was observed, leading to more complex patterns than the pure modes of the surface harmonics.
At high excitation frequencies and larger viscosities of the electrolyte, the instability exhibits boundary-independent patterns, especially the classical hexagonal pattern reported for mechanical actuation \citep{Edwards1994}.

With respect to the relevance of these results, at least two aspects deserve to be mentioned. 
First, we were able to clarify some of the observations made by \citet{Ward2019}. Specifically, it was reported that always a multitude of wavelengths emerged for all unstable configurations, and that no discrete modes were observed. These results contrast observations made for mechanically induced Faraday instabilities, as well as the results presented in our work. It is likely that in the aforementioned work, the discrete modes were obscured by edge-waves emerging from the meniscus at the domain boundary. 
Second, our results indicate that other phenomena, which were so far only observed for the mechanically induced instability, should as well emerge for electrostatic forcing. For example, in large domains the formation of superlattices and quasi-patterns upon multi-frequency actuation is expected \citep{Edwards1994}. Also, our work may pave the way to novel explorations of other effects so far only explored in the context of mechanical actuation, such as the walking-droplet phenomenon \citep{Couder2005, Bush2015, Fernandez-Mateo2021}.
Overall, this work draws far-reaching analogies to mechanically actuated Faraday waves, but may enable different schemes of studying such phenomena, due to the fundamentally different actuation mechanism.

\backsection[Acknowledgements]{The authors thank Julian Keller and Jörg Bültemann for designing an initial version of the setup and the technical support.}

\backsection[Funding]{This work was supported by the Deutsche Forschungsgemeinschaft (S.D., M.H., S.H., Grant No. HA 2696/45-1) and the Council of Scientific and Industrial Research (A.B., grant no. 22(0843)/20/EMR-II).}

\backsection[Declaration of interests]{The authors report no conflict of interest.}

\backsection[Author ORCID]{S. Dehe, https://orcid.org/0000-0002-9206-9106; M. Hartmann, https://orcid.org/0000-0001-5586-8753; A. Bandopadhyay, https://orcid.org/0000-0003-3371-7879; S. Hardt, https://orcid.org/0000-0001-7476-1070}

\appendix

\section{Fluid Properties Measurement Techniques}
\label{sec:appA-substances}

In order to make theoretical predictions and to reconstruct the interface from the experimental data, information about the fluid properties is required.
The relevant properties are interfacial tensions, densities, viscosities, relative permittivity, and refractive indices.
The measurement techniques used to determine the values presented in table~\ref{tab:fluidDataExcIFT} and table~\ref{tab:IFTs} are briefly described in the following.
The experiments were performed at a constant laboratory temperature of \SI{20}{\celsius}.\\

\textbf{Interfacial tension:}
The interfacial tensions between all combinations of bottom and top phase were determined with a ring tensiometer (\textit{DCAT 25}, \textit{DataPhysics Instruments}, Germany) with the specific ring \textit{RG 11}.
For each liquid combination, the interfacial tension was measured five times in total using the software delivered with the tensiometer.
Following the recommendations of the manual, the first measurement was performed using the \textit{pull 2 step} method as implemented in the manufacturer's software, while the subsequent four measurements were performed using the \textit{push/pull lamella} method. \\
%This is possible, since bottom and top liquid are highly immiscible.
%In both cases, the maximum detection of the weight measurement was performed.\\

\textbf{Dynamic viscosity:}
In case of the glycerol-water mixtures, the viscosity was determined using a digital rotation rheometer (\textit{DV-III-Ultra}, \textit{Brookfield}, USA) with the spindle \textit{CPE-40} having a coefficient of $\gamma =$307.
By setting the angular velocity $\omega$ and measuring the torque $M$ (in \% of the maximum torque), the dynamic viscosity $\eta$ can be calculated via
\begin{equation}
\eta = \frac{\omega \gamma}{M},
\label{eq:viscMeas}
\end{equation}
according to the manual.
For each liquid, the measurement was repeated at least five times.\\

\textbf{Density:}
The density was measured by dosing at least five individual droplets of known volume of the relevant liquid on a precision scale (\textit{NewClassic MF / MS 105DU}, \textit{Mettler Toledo}, Switzerland) using a \SI{5}{\milli \liter} syringe (Injekt Luer Solo, \textit{Braun}, Germany).
%Since the volume is known and the mass of the droplet can be measured almost instantaneously after it is deposited on the surface, the density can be calculated.
The experiment was repeated at least five times per fluid mixture.\\

\textbf{Refractive Indices:}
Refractive indices were measured with an Abbe refractometer (\textit{Carl Zeiss}, Germany).

\section{Theoretical description}
\label{sec:appB-Theory}

The theoretical description of the instability follows the procedure described in \citet{Bandopadhyay2017}, with the additional assumption of a thin upper layer ($h_1/h_2 \ll 1$).
In the following, an outline of the derivation of the model is given.
We refer to the schematic shown in figure~\ref{fig:experimentalSetup}, where the thickness of the aqueous layer is denoted by $h_2$, while the thickness to the silicone oil layer on top of the aqueous solution is denoted by $h_1$.
We choose the coordinate system to be attached to the undeformed surface. Let $\mathbf{x}'_s$ represent the surface coordinates, i.e. coordinates in the plane of the undeformed surface, while $z'$ represent the normal coordinate. Primed variables represent dimensional quantities.
With the convention above, gravity acts in the $-z'$ direction.
In order to arrive at the theoretical description for the interface deformation driven by the time-periodic electric field, we first assume the presence of a deformed interface, whose shape is given by $z' = \zeta'(\mathbf{x}'_s, t')$.
For this interface, in the limit of small deformation the normal direction, $\mathbf{n}$, and curvature, $\kappa$, are given by
\begin{eqnarray}
\mathbf{n} = \frac{\nabla (\zp-\zetap(\xsp,\tp))}{|\nabla (\zp-\zetap(\xsp,\tp))|} \approx
\mathbf{e}_z - \nablap_s \zeta(\xsp,\tp), \qquad \kappa = \nabla\cdot \mathbf{n} =
-\nabla_s^2 \zeta(\xsp,\tp),
\end{eqnarray}

Since the lower fluid is conducting it will be an isopotential volume. Therefore, only the potential in the insulating top phase, $\phi'_1 (\mathbf{x}'_s, z', t')$ needs to be evaluated and is obtained from
\begin{eqnarray}
\nabla^2\phi'_1 = 0\quad \textrm{subject to} \quad \phi'_1(\mathbf{x}'_s, h_1, t') = -V'(t), \; \phi'_1(\mathbf{x}'_s, \zeta'(\mathbf{x}'_s,t'), t') = V'(t),
\end{eqnarray}
where $V'(t)$ represents the applied time periodic potential. The hydrodynamics is governed by conservation of mass and momentum, with the latter expressed as
\begin{equation}
   \rho_j \left( \frac{\partial }{\partial t} + \up_j\cdot\nablap \right) \up_j = -\nablap \Pp_j+\eta_j{\nablap}^2\up_j-\rho_jg{\mathbf{e}_z}, \qquad j = 1,2;
\end{equation}
subject to the boundary conditions
\begin{eqnarray}
  \text{at} \quad \zp = h_1^\prime &:& \quad w_1^\prime = 0, \quad {\mathbf{u_s}}_1^\prime = 0 \\
  \text{at} \quad \zp = -h_2^\prime &:& \quad w_2^\prime = 0, \quad {\mathbf{u_s}}_2^\prime = 0 \\
  \text{at} \quad \zp = \zetap(\xsp,\tp) &:&\quad  w_1^\prime = w_2^\prime = \frac{\partial \zetap}{\partial \tp} + \up_1\cdot \zetap(\xsp,\tp), \qquad {\mathbf{u_s}}_1^\prime = {\mathbf{u_s}}_2^\prime\\
\text{at}\quad \zp = \zetap(\xsp,\tp) &:& \quad  \eta_1\left(\frac{\partial \up_1}{\partial \zp} + \nablap_sw_1^\prime \right)
  =\eta_2 \left(\frac{\partial \up_2}{\partial \zp} + \nablap_sw_1^\prime \right)\\
  \text{at}\quad \zp = \zetap(\xsp,\tp) &:& \quad \left(-\Pp _1 + 2\eta_1 \frac{\partial w_1^\prime}{\partial \zp}\right) - \left(-\Pp_2 + 2\eta_2 \frac{\partial w_2^\prime}{\partial \zp}\right) + {\mathbf{n}}\cdot{\mathbf{\tau_E^\prime}}\cdot{\mathbf{n}} = \sigma \kappa.
\end{eqnarray}
We can render the equations dimensionless with the help of the following scheme 
\begin{equation}
\begin{gathered}
  (\xs,z,\zeta) = \frac{(\xsp,\zp,\zetap)}{h_1},\quad
  t = \tp \omega, \quad
  {\us}_j = \frac{{\up}_{sj}}{h_1\omega},\quad
  P = \frac{\Pp}{\eta_1 \omega},\quad
  h = \frac{h_2}{h_1},\quad \\
  \eta_r = \frac{\eta_2}{\eta_1}, \quad
  \rho_r = \frac{\rho_2}{\rho_1}, \quad
  V(t) = \frac{V^\prime}{V_\text{ref}}, \quad
  l_\text{vis} = \sqrt{\frac{\nu_1}{\omega}}, \quad \\
  \Re = \frac{h_1^2}{l_\text{vis}^2}, \quad
  \Ma = \frac{\epsilon_1 V_\text{ref}^2/h_1^2}{\eta_1 \omega}, \quad 
  \Ca = \frac{\eta_1h_1\omega}{\sigma},\quad
  \Ga = \frac{\rho_1 g}{\eta_1\omega/h_1},
\end{gathered}
\end{equation}
where $\omega$ is the applied frequency, $l_{vis}$ represents the viscous lengthscale, and $\Re$ represents the Reynolds number, which may also be interpreted as the ratio of the characteristic length to the viscous lengthscale squared.
$\Ma$ represents the Mason number essentially quantifying the ratio of the normal Maxwell stress to the viscous stress.
$\Ca$ represents the capillary number and $\Ga$ represents the Galileo number.
The scale of pressure is chosen to be the viscous scale.

With the aforementioned nondimensional numbers, the governing equation for the potential may be written as
\begin{equation}
  \nabla^2\phi_1 = 0, \qquad \text{subjected to},\quad \phi_1(\xs,1,t) = -V(t),\quad \text{and} \quad \phi_1(\xs,\zeta,t) = V(t).
  \label{eq:pot_nd}
  \end{equation}
The hydrodynamic equations for the two fluids are rewritten as
\begin{eqnarray}
  \begin{aligned}
 \left( \frac{\partial }{\partial t} + \vel_1\cdot \nabla \right)\vel_1 &= -\frac{1}{\Re}\nabla P_1 + \frac{1}{\Re}\nabla^2 \vel_1 -\frac{\Ga}{\Re} {\mathbf{e}_z}, \\
 {\rho_r} \left( \frac{\partial }{\partial t} + \vel_2 \cdot \nabla \right)\vel_2 &= -\frac{1}{\Re}\nabla P_2 + \eta_r\frac{1}{\Re}\nabla^2 \vel_2 -\frac{\Ga}{\Re}\rho_r {\mathbf{e}_z},
  \end{aligned}
  \label{eq:hydro}
\end{eqnarray}
with appropriate boundary conditions at the top and bottom walls and interface given by
\begin{align}
  \text{at} \: z = 1: &\quad w_1 = 0, \quad \frac{\partial w_1}{\partial z} = 0
  \label{eq:noslip_top}
  \\
  \text{at} \quad z = -h: &\quad  w_2 = 0, \quad \frac{\partial w_2}{\partial z} = 0
  \label{eq:noslip_bot}
  \\
  \text{at} \: z = \zeta(\xs,t): &\quad  w_1 = w_2 = \frac{\partial \zeta}{\partial t} + {\us}_1\cdot \zeta(\xs,t), \qquad \frac{\partial w_1}{\partial z} = \frac{\partial w_2}{\partial z}
  \label{eq:dyn_bc}
  \\
\text{at} \: z = \zeta(\xs,t): &\quad \left(\frac{\partial {\us}_1}{\partial z} + \nabla_sw_1 \right)
  =\eta_r \left(\frac{\partial {\us}_2}{\partial z} + \nabla_sw_1 \right)
  \label{eq:tang_str}
  \\
  \text{at} \: z = \zeta(\xs,t): &\: \left(-P_1 + 2 \frac{\partial w_1}{\partial z}\right) - \left(-P_2 + 2\eta_r \frac{\partial w_2}{\partial z}\right) + \Ma\left({\mathbf{n}}\cdot{\mathbf{\tau_E}}\cdot{\mathbf{n}}\right) = -\frac{1}{\Ca}\nabla_s^2\zeta.
    \label{eq:mason}
\end{align}
The equations above may be solved analytically by a domain perturbation method described in \citet{Bandopadhyay2017}.
Finally, the governing equation relating the $z$ velocity and the interfacial deformation is given by
\begin{multline}
  \nabla_s^2\left(2(1-\eta_r)\frac{\partial w_1}{\partial z}+\frac{1}{\Ca}\nabla_s^2\zeta + \Ga \zeta (1-\rho_r) + 4\Ma\zeta k \coth k V(t)^2 \right) \\= \Re \left(\frac{\partial^2 w_1}{\partial z\partial t} -\rho_r \frac{\partial^2 w_2}{\partial z\partial t}\right) + \nabla^2\left( -\frac{\partial w_1}{\partial z} + \eta_r\frac{\partial w_2}{\partial z}\right).
  \label{eq:mason2}
\end{multline}
In order to make analytical progress, the aforementioned quantities may be written as
\begin{equation}
(w_i,\zeta) = (\hat{w}_i(z,t), \zc) \sin \bk\cdot\xs,
\end{equation}
where the terms are essentially split into time-dependent out-of-plane components and in-plane components.

In order to assess the stability of the interface under the applied electric field, we may represent the out-of-plane components as being composed of a time-dependent amplitude term and a time-periodic term; the former quantifying whether the amplitude grows exponentially in time.
\begin{gather}
\hat{w}_j(z,t) = \exp(st+i\alpha t)W_j(z,t\; \text{mod} (2\pi)) =  
\exp(st+i\alpha t) \sum\limits_{n=-\infty}^{n=\infty} W_{j,n}(z) \exp(in t),\\
\zc = \exp(st+i\alpha t) \sum\limits_{n=-\infty}^{n=\infty} Z_{n}(z)\exp(in t).
\label{eq:var_forms}
\end{gather}
In case it does not grow exponentially in time, the system can be classified as being stable.
Therefore, the behaviour of the parameter $s$ determines the stability of the system, while $\alpha$ governs the time-periodic nature of the system. If the real part of $s$ is positive, then the system is unstable, while negative real parts indicate a stable system. If $s$ is imaginary, then the solution remains periodic.
Accordingly, the system's stability limit may be probed by setting $s =0$ and finding the corresponding points in the wavenumber - Mason-number space \citep{Bandopadhyay2017}. The Floquet multiplier $\exp(st+i\alpha t)$ is nonunique since $s+2\pi j$ ($j$ being an integer) is also a solution. The constraint $0\leq \alpha \leq 1/2$ makes it unique, where $\alpha = 0$ represents the harmonic response of the system and $\alpha = 1/2$ the sub-harmonic response. Any integer increments to $\alpha$ may be absorbed in the sum and therefore do not yield any additional information.

Substituting this form of $w_i$ and $\zeta$ in the governing equation, we obtain
\begin{eqnarray}
\left( s+i(\alpha + n) - \frac{1}{\Re}\left( \frac{d^2 }{dz^2}- k^2\right)
\right) \left( \frac{d^2 }{dz^2}-k^2 \right) W_{1,n} = 0,\\
\left( s+i(\alpha + n) - \frac{\eta_r}{\rho_r\Re}\left( \frac{d^2 }{dz^2}- k^2\right)
\right) \left( \frac{d^2 }{dz^2}-k^2 \right) W_{2,n} = 0.
\label{eq:govern_velo_modes}
\end{eqnarray}
The expressions may be simplified using the abbreviations
%===============================
\begin{eqnarray}
  q_n^2 = k^2 + \Re (s + i (\alpha + n)),\qquad
  r_n^2 = k^2 + \Re\frac{\rho_r}{\eta_r} (s + i (\alpha + n)).
  \label{eq:q_r}
\end{eqnarray}
%===============================
Utilizing this, we obtain the solution for the vertical velocity components as
%===============================
\begin{eqnarray}
W_{1,n} &= P_{1,n} \exp (kz) + Q_{1,n}\exp (-kz) + R_{1,n}\exp (q_n z) + S_{1,n}
\exp (-q_n z) \\
W_{2,n} &= P_{2,n} \exp (kz) + Q_{2,n}\exp (-kz) + R_{2,n}\exp (r_n z) + S_{2,n}
\exp (-r_n z).
\label{eq:W_comp_sol}
\end{eqnarray}
The constants may be evaluated using the boundary conditions listed above as 
\begin{eqnarray}
\text{at} \: z = 1: &\quad&  W_{1,n} = 0, \;\; \frac{d W_{1,n}}{d z}(1) = 0, \\
  \label{eq:bcs_10}
\text{at} \: z = -h: &\quad& W_{2,n}(-h) = 0, \;\; \frac{d W_{2,n}}{d z}(-h) = 0, \\
\text{at} \: z = \zeta(\xs,t): &\quad& W_{1,n}(0) = W_{2,n}(0),\;\; \frac{d W_{1,n}}{d z}(0) = \frac{d W_{2,n}}{d z}(0), \\
 &\quad& \left( q_n^2 - k^2 \right) \frac{1}{\Re} Z_n = W_{1,n},\\
 &\quad& \left( k^2 + \frac{d^2 }{d z^2} \right)W_{1,n} = \eta_r\left( k^2 + \frac{d^2 }{d z^2} \right)W_{2,n}.
\label{eq:bcs_1}
\end{eqnarray}

Note that in the experiments of this work, the thickness of the lower liquid layer is significantly larger than the viscous lengthscale, implying that the dimensionless parameter $h_2 \ll 1$. 
For the present setup, $h_2\ll 1$ implies that we may effectively model the bottom layer as being infinitely deep, thereby obviating the need for boundary conditions at the lower wall, eliminating terms involving $\exp(-kz)$ in the above equations. In such a case, the solution to equations (\ref{eq:bcs_10} to \ref{eq:bcs_1}) may be written as
\begin{eqnarray}
W_{1,n} &=& P_{1,n} \exp (kz) + Q_{1,n}\exp (-kz) + R_{1,n}\exp (q_n z) + S_{1,n}
\exp (-q_n z) \\
W_{2,n} &=& P_{2,n} \exp (kz)+ R_{2,n}\exp (r_n z).
\label{eq:new_sol}
\end{eqnarray}

We can now evaluate the normal stress balanace, equation \eqref{eq:mason2}, by utilizing the Floquet form of the coefficients, equation \eqref{eq:var_forms}.
In equation \eqref{eq:var_forms}, the nonlinearity in the normal stress balance stems from the $V(t)^2$ term in $4\Ma \zeta k \coth k \, V(t)^2$.
Disregarding the prefactor, we can write the normal stress in the form
\begin{equation}
\zeta V(t)^2 = V(t)^2 \sum_{n=-\infty}^{n=\infty} \exp\left(s+i(\alpha + n)t \right)Z_n.
\end{equation}
In the expression above, we can make use of the following simplification: The rescaled voltage, $V(t)$ is represented in general as $V(t) = \sum_{n=-\infty}^{n=\infty} \beta_n \cos(nt) + \sum_{n=-\infty}^{n=\infty}\gamma_n \sin(nt)$.
The various harmonic modes appearing when evaluating $V(t)^2$ then couple with the $\exp(nt)$ mode of the $Z_n$ term above.
Substituting this, we find 
\begin{equation}
  \begin{split}
  \left( q_n^2 - k^2\right) \frac{d W_{1,n}}{dz}
  -\eta_r\left( r_n^2 - k^2\right) \frac{d W_{2,n}}{dz}
  -\left( \frac{d^2}{dz^2} - k^2\right)\frac{dW_{1,n}}{dz}
\\  +\eta_r\left(\frac{d^2}{dz^2} -k^2\right)\frac{dW_{2,n}}{dz}
  +2k^2\left(1-\eta_r \right)\frac{dW_{1,n}}{dz}
\\  =
  \frac{1}{\Ca}k^4Z_n
  -\Ga k^2\left(1-\rho_r \right)Z_n
  -4\Ma k^3\coth k\,V(t)^2 \zeta
  \end{split}
  \label{eq:norm_str_fin}
  \end{equation}
These equations are finally cast in the form of a generalized eigenvalue problem
\begin{equation}
\matr{A} \vec{Z} = \textrm{Ma} \matr{B} \vec{Z}.
\end{equation}
Examples of how the matrices on the left-hand and right-hand side look like are given in \citet{Bandopadhyay2017}.

\section{Data Evaluation}
\label{sec:appC-eval}

\subsection{Measurement of the surface gradient $\nabla h(x,y)$}
As a first step, the corner-points of the grid are extracted from the video. For this purpose, the grey-scale of the images is inverted, such that the grid appears as bright lines on a dark background. Next, the effect of non-uniform illumination is accounted for by subtracting the local mean value computed in a $\SI{60}{px} \times \SI{60}{px}$ region around each image point.
In the following, the image is further processed using the morphological closing and erosion operations of the \textit{Python} toolbox \textit{skimage.morphology} \citep{van2014scikit}, in order to reduce the thickness of each line in the image.
On this pre-processed image, a Harris corner-detection algorithm implemented in the Python-toolbox \textit{Skimage} is used to generate an initial guess for each corner point. Then, based on these initial positions, the subpixel position of each corner is obtained using a statistics-based algorithm \citep{forstner1987}.
As a result, we obtain the positions of the grid lines intersections. For an unperturbed interface, they appear \SI{0.5}{\milli \meter} apart.

As a next step, the grid points are tracked in time. In order to balance accuracy and computational cost, the points are tracked over one period of oscillation. Since the videos are recorded at a constant frame rate of \SI{1000}{fps}, the number of frames corresponding to one period varies, depending on the driving frequency. Also, for lower frequencies, we reduce the number of frames (using every second frame for $\SI{4}{\hertz}-\SI{7}{\hertz}$ driving frequency, every third for $\SI{3}{\hertz}$, and every fifth for $\SI{2}{\hertz}$). Nevertheless, each oscillation period is resolved using at least 50 frames. The grid points are tracked between frames based on a nearest neighbor method, with a maximum displacement between frames of \SI{5}{px}. As a result, the time evolution of the grid is obtained, where $\vec{x}_i(t)$ denotes the position of the \textit{i}-th grid point at time $t$. 

We then calculate the displacement field $\Delta \vec{x}_i\left( t \right)$ of each point $i$ using
\begin{equation}
\Delta \vec{x}_i\left( t \right) = \vec{x}_i\left( t \right) - \frac{1}{N} \sum_{t =1}^N \vec{x}_i, 
\end{equation}
where $N$ denotes the number of images within one period of oscillation.
Due to refraction, the image of the evenly spaced grid exhibits an uneven spacing between grid points, leading to an uneven spacing of the displacement field data. In preparation for the next step, the deviation field is interpolated onto a regular grid ($M\times M = 51 \times 51$ grid points, \SI{20}{px} spacing between points) using a spline interpolation. Finally, utilizing equations \eqref{eq:ExpDetails_h_gradient} and \eqref{eq:ExpDetails_hstar}, we obtain the surface gradient field $\nabla {h}(x,y)$ as a function of time \citep{Moisy2009}.

\subsection{Reconstruction of the interface shape $h(x,y)$}

From the surface gradient $\nabla h(x,y)$, the interface shape $h(x,y)$ can be reconstructed using the numerical approach outlined by \citet{Moisy2009}, which is described in the following.
A discrete representation of $h$ at the grid points ($M\times M = 51 \times 51$) is of interest. Since the grid points are evenly spaced, we can express the $x$- and the $y$-component of the gradient $\nabla h$ in terms of $h$ by using a central difference approximation. For example, if the index $j$ denotes the $j$-th grid point in $x$-direction, and the index $l$ denotes the $l$-th grid point in $y$-direction, the gradient at position $j$,$l$ can be approximated as 
%\begin{equation}
\begin{align}
\frac{\partial h (j,l)}{\partial x}  &= \frac{h(j+1,l) - h(j-1,l)}{2 L} \label{eq:ExpDetails_h_centereddiff_x}\\
\frac{\partial h (j,l)}{\partial y}  &= \frac{h(j,l+1) - h(j,l-1)}{2 L}\label{eq:ExpDetails_h_centereddiff_y}, 
\end{align}
%\end{equation}
where $L$ denotes the spacing between two adjacent grid points in physical coordinates.
The gradients at the image boundary can be treated appropriately by using the forward and backward difference scheme. As a result, two linear systems of equations emerge, describing the gradients in $x$- and $y$-direction as $\matr{G}_x \vec{h} = \vec{p}_x$ and  $\matr{G}_y \vec{h} = \vec{p}_y$. Here, $\vec{h}$ contains the layer thickness $h$ at the grid points, sorted into a vector of dimension $(M^2,1)$. The vectors $\vec{p}_x$, $\vec{p}_y$ are of dimension $(M^2,1)$ and contain the components of $\nabla h$ in $x$- and $y$-direction at the grid points, and the matrices $\matr{G}_x $, $\matr{G}_y$ of dimension $(M^2,M^2)$ contain the coefficients of the finite difference approximation schemes. Both systems can be combined into one linear system as 
\begin{equation}
\label{eq:ExpDetails_h_fullproblem}
\begin{pmatrix} \matr{G}_x \\ \matr{G}_y \end{pmatrix} 
\cdot \vec{h} = 
\begin{pmatrix} \vec{p}_x \\ \vec{p}_y \end{pmatrix}.
\end{equation}
As a result, we arrive at $2M^2$ equations for $M^2$ points. Such an over-determined system can be solved as an minimization problem of the 
functional $\vert (\matr{G}_x, \matr{G}_y)^\text{T} \vec{h} - ( \vec{p}_x ,\vec{p}_y)^\text{T} \vert^2$ using an optimization algorithm (e.g., \textit{lsq\_linear} of the Python toolbox \textit{scipy}).
Since an optimization algorithm is used for the problem \eqref{eq:ExpDetails_h_fullproblem} instead of direct numerical integration of the obtained surface gradients $\nabla h$, the interface reconstruction is less prone to local errors. While in a direct integration approach the local errors accumulate, the optimization approach minimizes a global error.
Lastly, the resulting field $h(x,y)$ is interpolated onto the original image of dimension $N \times N = \SI{1024}{px} \times \SI{1024}{px}$, and the deviation from the unperturbed interface is denoted as $\Delta h(x,y) = h(x,y) - h_1$. A reconstructed interface displacement $\Delta h(x,y)$ is shown in Fig. \ref{fig:experimentalEval}(b) as an example.

\subsection{Dominant pattern wavelength $\lambda$}
In order to determine the dominant pattern wavelength $\lambda$, a discrete Fourier transform of the reconstructed interface shape $h(x,y)$ is used.
First, we place the computed displacement $\Delta h$ on $N \times N$ grid points at the center of a larger $3N \times 3N$ image, with the other values set to zero. The indices of the corresponding matrix are denoted $j$ and $l$. Then, we compute the two-dimensional discrete Fourier transform ${H}$ as
\begin{equation}
H(k_x,k_y) = \sum_{j=0}^{3N-1} \sum_{l=0}^{3N-1} \Delta h(j,l) e^{-i 2 \pi \left( \frac{j\,k_x}{3N} + \frac{l\,k_y}{3N} \right) }, 
\end{equation}
where $k_x$, $k_y$ denote the wavenumbers in $x$- and $y$-direction, respectively.
Since we are interested in the dominant wavelengths, independent of the spatial orientation of the waves, we compute the power spectrum of $H(k_x,k_y)$. The amplitude for wavenumbers $k_x$, $k_y$ is obtained as $A(k_x,k_y) = \vert H(k_x,k_y) \vert^2$ and the wavenumber magnitude as $k =  (k_x^2 +k_y^2)^{0.5}$.
Now, we bin the amplitudes $A(k_x,k_y)$ into $k$-values using bins ranging from 0 to $3N/2$, with a bin-size of $\text{d}k_\text{bin}=\SI{1}{px}$. Then, we normalize the resulting values inside each bin, using the corresponding bin area of the Fourier transform $\pi \left(k_\text{bin}+\text{d}k_\text{bin} \right)^2 - \pi  k_\text{bin}^2 $, where $k_\text{bin}$ denotes the lower boundary of the bin. As a result, the power spectrum of the interface deformation is obtained in terms of $k$.  

The resulting power spectrum yields information about the dominant wavelengths of the system, and shares some distinct properties between all cases considered. An exemplary power spectrum is shown in Fig. \ref{fig:experimentalEval}(c). First, at small wavenumbers (corresponding to large wavelengths), a distinct peak is present, denoted by (I). This is a result of embedding the original image matrix in a domain of dimension $3N \times 3N$. Also, at large wavenumbers (corresponding to small wavelengths), another distinct peak is present, denoted by (III). This results from the reconstruction of the interface using a reduced grid with a spacing of \SI{20}{px} between grid points. Both of these peaks are artifacts without physical interpretation. A third peak can be observed in between these two if the interface shows a distinct pattern. The wavenumber $k_\text{max}$ of this peak corresponds to the dominant wavelength $\lambda$ of the system, which is obtained as
\begin{equation}
\lambda = (3N) a_\text{cal} / k_\text{max}.
\end{equation}
For each image of the series of experiments, the peaks in the power spectrum are detected using the Python toolbox \textit{scipy} (function \textit{signal.find\_peaks}).
In order to enhance the accuracy of the wavelength detection from the power spectrum with discrete wave numbers $k$, we fit a Gaussian distribution to the portion of the power spectrum closest to the detected peak as initial guess. Then, we identify the determined mean value as the dominant wavenumber that serves as an input for further calculations.
Depending on the nature of the oscillations, at some of the time steps, no peaks are detected in the power spectrum.
Ideally, the spectrum exhibits only one dominant wavelength. However, due to waves originating from the boundary of the container, different wavelengths can be present in one image, with each wavelength corresponding to either the Faraday waves or the edge waves. 
In order to accurately determine the wavelengths, the histogram of wavelengths over all images is obtained, exhibiting a superposition of multiple peaks. Following, the average value of each distinct peak of the histogram is computed, and the multiple wavelengths occurring in one experiment can be retrieved. In the case of multiple wavelengths detected in one experiment, the wavelength observed in the highest number of images is denoted as \textit{dominant}, and other wavelengths are denoted as \textit{secondary}. The attributed error bars denote the standard deviation of the pattern wavelengths obtained during one experiment and characterizes the error attributed to the pattern detection.

\bibliographystyle{jfm}
%\bibliography{literature}

\begin{thebibliography}{49}
\expandafter\ifx\csname natexlab\endcsname\relax\def\natexlab#1{#1}\fi
\def\au#1{#1} \def\ed#1{#1} \def\yr#1{#1}\def\at#1{#1}\def\jt#1{\textit{#1}}
  \def\bt#1{#1}\def\bvol#1{\textbf{#1}} \def\vol#1{#1} \def\pg#1{#1}
  \def\publ#1{#1}\def\arxiv#1{#1}\def\org#1{#1}\def\st#1{\textit{#1}}

\bibitem[Bandopadhyay \& Hardt(2017)]{Bandopadhyay2017}
{\sc \au{Bandopadhyay, A.} \& \au{Hardt, S.}} \yr{2017}  \at{{Stability of
  horizontal viscous fluid layers in a vertical arbitrary time periodic
  electric field}}.  \jt{Physics of Fluids}  \bvol{29}~(12).

\bibitem[Batson {\em et~al.\/}(2013)Batson, Zoueshtiagh \&
  Narayanan]{Batson2013}
{\sc \au{Batson, W.}, \au{Zoueshtiagh, F.} \& \au{Narayanan, R.}} \yr{2013}
  \at{{Dual role of gravity on the Faraday threshold for immiscible viscous
  layers}}.  \jt{Physical Review E}  \bvol{88}~(6),  \pg{063002}.

\bibitem[Benjamin \& Ursell(1954)]{Benjamin1954}
{\sc \au{Benjamin, T.~B.} \& \au{Ursell, F.}} \yr{1954}  \at{{The stability of
  the plane free surface of a liquid in vertical periodic motion}}.
  \jt{Proceedings of the Royal Society of London. Series A. Mathematical and
  Physical Sciences}  \bvol{225}~(1163),  \pg{505--515}.

\bibitem[Besson {\em et~al.\/}(1996)Besson, Edwards \& Tuckerman]{Besson1996}
{\sc \au{Besson, T.}, \au{Edwards, W.~S.} \& \au{Tuckerman, L.~S.}} \yr{1996}
  \at{{Two-frequency parametric excitation of surface waves}}.  \jt{Physical
  Review E - Statistical Physics, Plasmas, Fluids, and Related
  Interdisciplinary Topics}  \bvol{54}~(1),  \pg{507--513}.

\bibitem[Briskman \& Shaidurov(1968)]{Briskman1968}
{\sc \au{Briskman, V.~A.} \& \au{Shaidurov, G.~F.}} \yr{1968}  \at{{Parametric
  Instability of a Fluid Surface in an Alternating Electric Field}}.
  \jt{Soviet Physics Doklady}  \bvol{13},  \pg{540}.

\bibitem[Bush(2015)]{Bush2015}
{\sc \au{Bush, John~W.M.}} \yr{2015}  \at{{Pilot-Wave Hydrodynamics}}.
  \jt{Annual Review of Fluid Mechanics}  \bvol{47}~(1),  \pg{269--292}.

\bibitem[Ciliberto \& Gollub(1984)]{Ciliberto1984}
{\sc \au{Ciliberto, S.} \& \au{Gollub, J.~P.}} \yr{1984}  \at{{Pattern
  Competition Leads to Chaos}}.  \jt{Physical Review Letters}  \bvol{52}~(11),
  \pg{922--925}.

\bibitem[Couder {\em et~al.\/}(2005)Couder, Proti{\`{e}}re, Fort \&
  Boudaoud]{Couder2005}
{\sc \au{Couder, Y.}, \au{Proti{\`{e}}re, S.}, \au{Fort, E.} \& \au{Boudaoud,
  A.}} \yr{2005}  \at{{Walking and orbiting droplets}}.  \jt{Nature}
  \bvol{437}~(7056),  \pg{208--208}.

\bibitem[Dodge {\em et~al.\/}(1965)Dodge, Kana \& Abramson]{Dodge1965}
{\sc \au{Dodge, F.~T.}, \au{Kana, D.~D.} \& \au{Abramson, H.~N.}} \yr{1965}
  \at{{Liquid surface oscillations in longitudinally excited rigid cylindrical
  containers}}.  \jt{AIAA Journal}  \bvol{3}~(4),  \pg{685--695}.

\bibitem[Douady(1990)]{Douady1990}
{\sc \au{Douady, S.}} \yr{1990}  \at{{Experimental study of the Faraday
  instability}}.  \jt{Journal of Fluid Mechanics}  \bvol{221}~(5),
  \pg{383--409}.

\bibitem[Douady \& Fauve(1988)]{Douady1988}
{\sc \au{Douady, S.} \& \au{Fauve, S.}} \yr{1988}  \at{{Pattern selection in
  faraday instability}}.  \jt{Epl}  \bvol{6}~(3),  \pg{221--226}.

\bibitem[Edwards \& Fauve(1994)]{Edwards1994}
{\sc \au{Edwards, W.~S.} \& \au{Fauve, S.}} \yr{1994}  \at{{Patterns and
  quasi-patterns in the Faraday experiment}}.  \jt{Journal of Fluid Mechanics}
  \bvol{278}~(II),  \pg{123--148}.

\bibitem[Epstein \& Fineberg(2008)]{Epstein2008}
{\sc \au{Epstein, T.} \& \au{Fineberg, J.}} \yr{2008}  \at{{Necessary
  Conditions for Mode Interactions in Parametrically Excited Waves}}.
  \jt{Physical Review Letters}  \bvol{100}~(13),  \pg{134101}.

\bibitem[Faraday(1831)]{Faraday1831}
{\sc \au{Faraday, M.}} \yr{1831}  \at{{XVII. On a peculiar class of acoustical
  figures; and on certain forms assumed by groups of particles upon vibrating
  elastic surfaces}}.  \jt{Philosophical Transactions of the Royal Society of
  London}  \bvol{121}~(August 1827),  \pg{299--340}.

\bibitem[Fern{\'{a}}ndez-Mateo \& P{\'{e}}rez(2021)]{Fernandez-Mateo2021}
{\sc \au{Fern{\'{a}}ndez-Mateo, R.} \& \au{P{\'{e}}rez, A.~T.}} \yr{2021}
  \at{{Faraday waves under perpendicular electric field and their application
  to the walking droplet phenomenon}}.  \jt{Physics of Fluids}  \bvol{33}~(1),
  \pg{017109}.

\bibitem[F{\"o}rstner \& G{\"u}lch(1987)]{forstner1987}
{\sc \au{F{\"o}rstner, W.} \& \au{G{\"u}lch, E.}} \yr{1987} A fast operator for
  detection and precise location of distinct points, corners and centres of
  circular features.  \bt{In {\em Proc. ISPRS intercommission conference on
  fast processing of photogrammetric data\/}},  \pg{pp. 281--305}. Interlaken.

\bibitem[Gambhire \& Thaokar(2014)]{Gambhire2014}
{\sc \au{Gambhire, P.} \& \au{Thaokar, R.}} \yr{2014}  \at{{Electrokinetic
  model for electric-field-induced interfacial instabilities}}.  \jt{Physical
  Review E}  \bvol{89}~(3),  \pg{032409}.

\bibitem[Gambhire \& Thaokar(2010)]{Gambhire2010}
{\sc \au{Gambhire, P.} \& \au{Thaokar, R.~M.}} \yr{2010}
  \at{{Electrohydrodynamic instabilities at interfaces subjected to alternating
  electric field}}.  \jt{Physics of Fluids}  \bvol{22}~(6),  \pg{064103}.

\bibitem[Gambhire \& Thaokar(2012)]{Gambhire2012}
{\sc \au{Gambhire, P.} \& \au{Thaokar, R.~M.}} \yr{2012}  \at{{Role of
  conductivity in the electrohydrodynamic patterning of air-liquid
  interfaces}}.  \jt{Physical Review E}  \bvol{86}~(3),  \pg{036301}.

\bibitem[Gluckman {\em et~al.\/}(1993)Gluckman, Marcq, Bridger \&
  Gollub]{Gluckman1993}
{\sc \au{Gluckman, B.~J.}, \au{Marcq, P.}, \au{Bridger, J.} \& \au{Gollub,
  J.~P.}} \yr{1993}  \at{{Time averaging of chaotic spatiotemporal wave
  patterns}}.  \jt{Physical Review Letters}  \bvol{71}~(13),  \pg{2034--2037}.

\bibitem[Gollub \& Meyer(1983)]{Gollub1983}
{\sc \au{Gollub, J.~P.} \& \au{Meyer, C.~W.}} \yr{1983}  \at{{Symmetry-breaking
  instabilities on a fluid surface}}.  \jt{Physica D: Nonlinear Phenomena}
  \bvol{6}~(3),  \pg{337--346}.

\bibitem[Iino {\em et~al.\/}(1985)Iino, Suzuki \& Ikushima]{Iino1985}
{\sc \au{Iino, M.}, \au{Suzuki, M.} \& \au{Ikushima, A.~J.}} \yr{1985}
  \at{{Surface-Wave Resonance Method for Measuring Surface Tension With a Very
  High Precision}}.  \jt{Le Journal de Physique Colloques}  \bvol{46}~(C10),
  \pg{C10--813--C10--816}.

\bibitem[Jones \& Melcher(1973)]{Jones1973}
{\sc \au{Jones, T.~B.} \& \au{Melcher, J.~R.}} \yr{1973}  \at{{Dynamics of
  electromechanical flow structures}}.  \jt{Physics of Fluids}  \bvol{16}~(3),
  \pg{393}.

\bibitem[Kahouadji {\em et~al.\/}(2015)Kahouadji, P{\'{e}}rinet, Tuckerman,
  Shin, Chergui \& Juric]{Kahouadji2015}
{\sc \au{Kahouadji, L.}, \au{P{\'{e}}rinet, N.}, \au{Tuckerman, L.~S.},
  \au{Shin, S.}, \au{Chergui, J.} \& \au{Juric, D.}} \yr{2015}  \at{{Numerical
  simulation of supersquare patterns in Faraday waves}}.  \jt{Journal of Fluid
  Mechanics}  \bvol{772},  \pg{R2.1--R2.12}.

\bibitem[Kityk {\em et~al.\/}(2005)Kityk, Embs, Mekhonoshin \&
  Wagner]{Kityk2005}
{\sc \au{Kityk, A.~V.}, \au{Embs, J.}, \au{Mekhonoshin, V.~V.} \& \au{Wagner,
  C.}} \yr{2005}  \at{{Spatiotemporal characterization of interfacial Faraday
  waves by means of a light absorption technique}}.  \jt{Physical Review E}
  \bvol{72}~(3),  \pg{036209}.

\bibitem[Kumar \& Tuckerman(1994)]{Kumar1994}
{\sc \au{Kumar, K.} \& \au{Tuckerman, L.~S.}} \yr{1994}  \at{{Parametric
  Instability of the Interface Between two Fluids}}.  \jt{Journal of Fluid
  Mechanics}  \bvol{279}~(May 2014),  \pg{49--68}.

\bibitem[Matthiessen(1868)]{Matthiessen1868}
{\sc \au{Matthiessen, L.}} \yr{1868}  \at{{Akustische Versuche, die kleinsten
  Transversalwellen der Fl{\"{u}}ssigkeiten betreffend}}.  \jt{Annalen der
  Physik und Chemie}  \bvol{210}~(5),  \pg{107--117}.

\bibitem[Matthiessen(1870)]{Matthiessen1870}
{\sc \au{Matthiessen, L.}} \yr{1870}  \at{{Ueber die Transversalschwingungen
  t{\"{o}}nender tropfbarer und elastischer Fl{\"{u}}ssigkeiten}}.  \jt{Annalen
  der Physik und Chemie}  \bvol{217}~(11),  \pg{375--393}.

\bibitem[Melcher(1966)]{Melcher1966}
{\sc \au{Melcher, J.~R.}} \yr{1966}  \at{{Traveling-Wave Induced
  Electroconvection}}.  \jt{Physics of Fluids}  \bvol{9}~(8),  \pg{1548}.

\bibitem[Melcher \& Schwarz(1968)]{Melcher1968}
{\sc \au{Melcher, J.~R.} \& \au{Schwarz, W.~J.}} \yr{1968}  \at{{Interfacial
  Relaxation Overstability in a Tangential Electric Field}}.  \jt{Physics of
  Fluids}  \bvol{11}~(12),  \pg{2604}.

\bibitem[Miles(1990)]{Miles1990}
{\sc \au{Miles, J.}} \yr{1990}  \at{{Parametrically Forced Surface Waves}}.
  \jt{Annual Review of Fluid Mechanics}  \bvol{22}~(1),  \pg{143--165}.

\bibitem[Moisy {\em et~al.\/}(2009)Moisy, Rabaud \& Salsac]{Moisy2009}
{\sc \au{Moisy, F.}, \au{Rabaud, M.} \& \au{Salsac, K.}} \yr{2009}  \at{{A
  synthetic Schlieren method for the measurement of the topography of a liquid
  interface}}.  \jt{Experiments in Fluids}  \bvol{46}~(6),  \pg{1021--1036}.

\bibitem[M{\"{u}}ller {\em et~al.\/}(2011)M{\"{u}}ller, Friedrich \&
  Papathanassiou]{Muller2011}
{\sc \au{M{\"{u}}ller, H.~W.}, \au{Friedrich, R.} \& \au{Papathanassiou, D.}}
  \yr{2011}  \at{{Theoretical and Experimental Investigations of the Faraday
  Instability}}.  \bt{In {\em Evolution of Spontaneous Structures in
  Dissipative Continuous Systems\/}}, ,  \vol{vol.~44},  \pg{pp. 230--265}.
  \publ{Berlin, Heidelberg: Springer Berlin Heidelberg}.

\bibitem[Nevolin(1984)]{Nevolin1984}
{\sc \au{Nevolin, V.~G.}} \yr{1984}  \at{{Parametric excitation of surface
  waves}}.  \jt{Journal of Engineering Physics}  \bvol{47}~(6),
  \pg{1482--1494}.

\bibitem[Perlin \& Schultz(2000)]{Perlin2000}
{\sc \au{Perlin, M.} \& \au{Schultz, W.~W.}} \yr{2000}  \at{{Capillary Effects
  on Surface Waves}}.  \jt{Annual Review of Fluid Mechanics}  \bvol{32}~(1),
  \pg{241--274}.

\bibitem[Pillai \& Narayanan(2018)]{Pillai2018}
{\sc \au{Pillai, D.~S.} \& \au{Narayanan, R.}} \yr{2018}  \at{{Nonlinear
  dynamics of electrostatic Faraday instability in thin films}}.  \jt{Journal
  of Fluid Mechanics}  \bvol{855},  \pg{R4}.

\bibitem[Rayleigh(1883)]{Rayleigh1883}
{\sc \au{Rayleigh, Lord}} \yr{1883}  \at{{XXXIII. On maintained vibrations}}.
  \jt{The London, Edinburgh, and Dublin Philosophical Magazine and Journal of
  Science}  \bvol{15}~(94),  \pg{229--235}.

\bibitem[Roberts \& Kumar(2009)]{Roberts2009}
{\sc \au{Roberts, S.~A.} \& \au{Kumar, S.}} \yr{2009}  \at{{AC
  electrohydrodynamic instabilities in thin liquid films}}.  \jt{Journal of
  Fluid Mechanics}  \bvol{631},  \pg{255--279}.

\bibitem[Robinson {\em et~al.\/}(2000)Robinson, Bergougnou, Cairns, Castle \&
  Inculet]{Robinson2000}
{\sc \au{Robinson, J.~A.}, \au{Bergougnou, M.~A.}, \au{Cairns, W.~L.},
  \au{Castle, G. S.~P.} \& \au{Inculet, I.~I.}} \yr{2000}  \at{{Breakdown of
  air over a water surface stressed by a perpendicular alternating electric
  field, in the presence of a dielectric barrier}}.  \jt{IEEE Transactions on
  Industry Applications}  \bvol{36}~(1),  \pg{68--75}.

\bibitem[Robinson {\em et~al.\/}(2001)Robinson, Bergougnou, Castle \&
  Inculet]{Robinson2001}
{\sc \au{Robinson, J.~A.}, \au{Bergougnou, M.~A.}, \au{Castle, G. S.~P.} \&
  \au{Inculet, I.~I.}} \yr{2001}  \at{{The electric field at a water surface
  stressed by an AC voltage}}.  \jt{IEEE Transactions on Industry Applications}
   \bvol{37}~(3),  \pg{735--742}.

\bibitem[Robinson {\em et~al.\/}(2002)Robinson, Bergougnou, Castle \&
  Inculet]{Robinson2002}
{\sc \au{Robinson, J.~A.}, \au{Bergougnou, M.~A.}, \au{Castle, G. S.~P.} \&
  \au{Inculet, I.~I.}} \yr{2002}  \at{{A nonlinear model of AC-field-induced
  parametric waves on a water surface}}.  \jt{IEEE Transactions on Industry
  Applications}  \bvol{38}~(2),  \pg{379--388}.

\bibitem[Serpooshan {\em et~al.\/}(2017)Serpooshan, Chen, Wu, Lee, Sharma, Hu,
  Venkatraman, Ganesan, Usta, Yarmush, Yang, Wu, Demirci \& Wu]{Chen2017}
{\sc \au{Serpooshan, V.}, \au{Chen, P.}, \au{Wu, H.}, \au{Lee, S.}, \au{Sharma,
  A.}, \au{Hu, D.~A.}, \au{Venkatraman, S.}, \au{Ganesan, A.~V.}, \au{Usta,
  O.~B.}, \au{Yarmush, M.}, \au{Yang, F.}, \au{Wu, J.~C.}, \au{Demirci, U.} \&
  \au{Wu, S.~M.}} \yr{2017}  \at{{Bioacoustic-enabled patterning of human
  iPSC-derived cardiomyocytes into 3D cardiac tissue}}.  \jt{Biomaterials}
  \bvol{131},  \pg{47--57}.

\bibitem[Shao {\em et~al.\/}(2021)Shao, Wilson, Saylor \& Bostwick]{Shao2021}
{\sc \au{Shao, X.}, \au{Wilson, P.}, \au{Saylor, J.~R.} \& \au{Bostwick,
  J.~B.}} \yr{2021}  \at{{Surface wave pattern formation in a cylindrical
  container}}.  \jt{Journal of Fluid Mechanics}  \bvol{915},  \pg{A19}.

\bibitem[Taylor \& McEwan(1965)]{Taylor1965}
{\sc \au{Taylor, G.~I.} \& \au{McEwan, A.~D.}} \yr{1965}  \at{{The stability of
  a horizontal fluid interface in a vertical electric field}}.  \jt{Journal of
  Fluid Mechanics}  \bvol{22}~(01),  \pg{1}.

\bibitem[Tufillaro {\em et~al.\/}(1989)Tufillaro, Ramshankar \&
  Gollub]{Tufillaro1989}
{\sc \au{Tufillaro, N.~B.}, \au{Ramshankar, R.} \& \au{Gollub, J.~P.}}
  \yr{1989}  \at{{Order-Disorder Transition in Capillary Ripples}}.
  \jt{Physical Review Letters}  \bvol{62}~(4),  \pg{422--425}.

\bibitem[Van~der Walt {\em et~al.\/}(2014)Van~der Walt, Sch{\"o}nberger,
  Nunez-Iglesias, Boulogne, Warner, Yager, Gouillart \& Yu]{van2014scikit}
{\sc \au{Van~der Walt, Stefan}, \au{Sch{\"o}nberger, Johannes~L},
  \au{Nunez-Iglesias, Juan}, \au{Boulogne, Fran{\c{c}}ois}, \au{Warner,
  Joshua~D}, \au{Yager, Neil}, \au{Gouillart, Emmanuelle} \& \au{Yu, Tony}}
  \yr{2014}  \at{scikit-image: image processing in python}.  \jt{PeerJ}
  \bvol{2},  \pg{e453}.

\bibitem[Ward {\em et~al.\/}(2019)Ward, Matsumoto \& Narayanan]{Ward2019}
{\sc \au{Ward, K.}, \au{Matsumoto, S.} \& \au{Narayanan, R.}} \yr{2019}
  \at{{The electrostatically forced Faraday instability: theory and
  experiments}}.  \jt{Journal of Fluid Mechanics}  \bvol{862},  \pg{696--731}.

\bibitem[Yih(1968)]{Yih1968}
{\sc \au{Yih, C.-S.}} \yr{1968}  \at{{Stability of a Horizontal Fluid Interface
  in a Periodic Vertical Electric Field}}.  \jt{Physics of Fluids}
  \bvol{11}~(7),  \pg{1447}.

\bibitem[Zhao {\em et~al.\/}(2019)Zhao, Dietzel \& Hardt]{Zhao2019}
{\sc \au{Zhao, S.}, \au{Dietzel, M.} \& \au{Hardt, S.}} \yr{2019}  \at{{Faraday
  instability of a liquid layer on a lubrication film}}.  \jt{Journal of Fluid
  Mechanics}  \bvol{879},  \pg{422--447}.

\end{thebibliography}

\newpage

\end{document}